\documentclass[11pt]{cernrep}
\usepackage{graphicx}
\begin{document}
\title{
$Z$-SCALING,  HIGH-$pT$ DIRECT PHOTON AND $ \pi^0$-MESON
 PRODUCTION AT RHIC AND LHC
% AT HIGH $P_T$
%Z-SCALING AND DIRECT PHOTON
% PRODUCTION
%% AT RHIC AND LHC}
% AT HIGH $P_T$
 }

\author{M.Tokarev, G.Ef{i}mov}
\institute{
Joint Institute for Nuclear Research,
141980, Dubna, Moscow region, Russia}
\maketitle
\begin{abstract}
The scaling properties  of direct  photon production in $pp, \bar pp$  and $pA$ collisions at high energies is
reviewed. The experimental data on the cross sections obtained at ISR, SpS and Tevatron are  used in the
analysis. The properties of data $z$-presentation, the energy and angular independencies, the power law, and
$A$-dependence are discussed.  The use of $z$-scaling to search for new physics phenomena in hadron-hadron,
hadron-nucleus and nucleus-nucleus collisions is suggested. The violation of $z$-scaling characterized by the
change of the fractal dimension is considered as a new and complimentary signature of a nuclear phase
transition. The  cross sections of direct photon, $\pi^0$- and $\eta^0$-meson production in $pp$ and $pPb$
collisions at RHIC and LHC energies are predicted.
\end{abstract}

%\newpage

{\section{INTRODUCTION}}
%A search for general properties of quark and gluon interactions
%beyond Quantum Chromodynamics (QCD)  in hadron-hadron,
%hadron-nucleus  and nucleus-nucleus collisions is one of the main
%goals of high energy particle and relativistic nuclear physics at
%RHIC and LHC. A high colliding energy allows us to study
%hadron-hadron  collisions in the framework of the perturbative
%QCD  and there is hopefulness to extract direct information on
%fundamental gluon distributions in the proton. It is considered
%that possible discrepancies with perturbative QCD predictions can
%be indicators of new  physics phenomena (quark compositeness, new
%types of interaction beyond the Standard Model etc.)

%Different signatures such as quarkonium suppression \cite{Satz},
%jet quenching \cite{Wang}, strangeness enhancement \cite{Raf} are
%used to search for indications of new physics phenomena such as a
%new state of nuclear matter named as Quark-Gluon Plasma (QGP)
%\cite{Shuryak,McLerran}.

Direct  photon  production due to small cross section of electromagnetic interactions is one among very few
signals which can provide  direct information on the partonic, early phase of interaction. 'Penetrating
probes', direct  photons   and dilepton pairs are traditionally   considered  to be one of the best probes for
Quark-Gluon Plasma (QGP) \cite{Feinberg,Shuryak,McLerran}.

Direct photons are produced through different mechanisms (Compton scattering, quark annihilation, bremsstrahlung
of quarks and gluons). Decay of hadrons such as $\pi^0$- and $\eta^0$-mesons is a source of background for direct
photon production. Therefore  it is important to establish  general features of photon production in hadron-hadron
($pp, \bar pp, \pi p$ etc.) collisions and then to search for their violation at extremal conditions (high
multiplicity particle density, high $p_T$ etc.) taking place in hadron-nucleus and nucleus-nucleus collisions.

%The main features of direct photon production  observed in $pp$
%interactions allow us to study the influence of nuclear matter
%on  photon formation in $pA$ and $AA$ collisions, to obtain
%restrictions on mechanism of photon formation given by pQCD and
%extract of the gluon structure function. A comparison of cross
%sections of direct photons produced in hadron-hadron,
%hadron-nucleus and nucleus-nucleus gives us a detailed
%understanding of the underlying  physical phenomena due to the
%presence of nuclear matter.

The description of direct photon production in the framework of QCD \cite{Auren,Jalil,Jeon} and comparison with
available experimental data reveal numerous ambiguities . Some of them are theoretical and  connected with the
choice of the factorization scheme, renormalization, factorization and fragmentation scales and with
considerations of higher order QCD corrections. The other ones are relevant to consistency among different
experimental data sets. The third ones are phenomenological and are introduced to correct theory by the model
dependent manner (for example "$k_T$"-smearing effect).

Nuclear effects such as multiple parton interactions, nuclear
shadowing , energy loss are not small \cite{Jalil} and  should be
taken into account for calculations of direct photon cross section
in heavy ion collisions. These effects modify nuclear structure
functions, photon fragmentation functions and add uncertainties
in the theoretical calculations of cross sections.

We use the concept of $z$-scaling to analyze numerous experimental data on direct photon cross sections for $pp$,
$\bar pp$ and $pA$ collisions at high  $ p_T$ and to make some predictions for $\gamma, \pi^0$ and $\eta^0$
produced in $pp$ and $pA$ collisions at the RHIC and LHC energies. The method of data analysis is complementary to
a method of direct calculations developed in the framework of QCD and methods based on Monte Carlo generators. The
use of the method allow us to obtain additional constraints in order to reduce the theoretical uncertainties and
to estimate more reliably the photon cross section and background.

\vskip 1cm

{\section{$Z$-SCALING}}

The idea of $z$-scaling \cite{Z96}-\cite{Z00} is based on the assumptions that inclusive particle distribution
of the process (\ref{eq:r1}) at high energies and high $p_T$ can be described in terms of the corresponding
kinematic characteristics
\begin{equation}
P_{1}+P_{2} \rightarrow q + X \label{eq:r1}
\end{equation}
of the exclusive elementary sub-process \cite{Stavinsky}
%written in the symbolic form (\ref{eq:r2})
%  \begin{equation}
%  (x_{1}M_{1}) + (x_{2}M_{2}) \rightarrow m_{1} +
%  (x_{1}M_{1}+x_{2}M_{2} + m_{2})
%  \label{eq:r2}
%  \end{equation}
and that the scaling function depending on a single variable $z$
exists and can be expressed via the dynamic quantities, invariant
inclusive  cross section $Ed^3\sigma/{dq^{3}}$ of the process
(\ref{eq:r1}) and particle multiplicity density $\rho(s,\eta)$.

%The kinematic quantities of the process (\ref{eq:r1}) are
%$P_1,P_2,q$ and $M_{1},  M_{2}, m_1$,
%  the momenta and masses of the colliding objects (hadron, nuclei)
%  and inclusive particles, respectively.
%   The parameter $m_{2}$ is introduced  to satisfy the  internal conservation
%  laws (for isospin, baryon number,
%  and strangeness). The $x_{1}$ and $x_{2}$ are the scale-invariant
%  fractions of the incoming four-momenta
%$P_{1}$ and $P_{2}$. They determine the   minimum energy, which
%  is necessary for the production of the secondary particle with
%  the mass $m_1$ and the four-momentum $q$.

  The elementary parton-parton collision is considered  as a binary
  sub-process which satisfies the condition

  \begin{equation}
  (x_{1}P_{1} + x_{2}P_{2} - q)^{2} = (x_{1}M_{1} + x_{2}M_{2} +
  m_{2})^{2}.
  \label{eq:r5}
  \end{equation}
  The equation reflects minimum recoil mass hypothesis in the
  elementary sub-process. To connect kinematic and structural
  characteristics of the interaction, the coefficient
  $\Omega$ is introduced. It is chosen in the form
  \begin{equation}
  \Omega(x_1,x_2) = m(1-x_{1})^{\delta_1}(1-x_{2})^{\delta_2},
  \label{eq:r8}
  \end{equation}
  where $m$ is a mass constant and $\delta_1$ and $\delta_2$
  are factors relating to the fractal structure of
  the colliding objects \cite{Z99}.
  The fractions $x_{1}$ and
  $x_{2}$  are determined  to maximize the value of $\Omega(x_1,x_2)$,
  simultaneously fulfilling the condition (\ref{eq:r5})
  \begin{equation}
  {d\Omega(x_1,x_2)/ dx_1}|_{x_2=x_2(x_1)} = 0.
  \label{eq:r9}
  \end{equation}
  The variables
  $x_{1,2}$ are equal to unity along the phase space limit and
  cover the full phase space accessible at any  energy.

{\subsection{Scaling function $\psi(z)$ and variable $z$}}

The scaling function $\psi(z)$ is written in the form \cite{Z99}
 \begin{equation}
 \psi(z) = - \frac{\pi s_A}{\rho_A(s,\eta) \sigma_{inel}}J^{-1}
 E\frac{d\sigma}{dq^{3}}.
 \label{eq:r20}
 \end{equation}
Here $\sigma_{inel}$ is the inelastic cross section, $s_A \simeq
s \cdot A$ and $s$ are the center-of-mass energy
 squared of the corresponding $ h-A $
 and $ h-N $ systems, $A$ is the atomic weight and
$\rho_A(s,\eta)$ is the average particle multiplicity density.
The factor $J$ is the known function of the kinematic variables,
the momenta and masses of the colliding and produced particles
\cite{Z99}.

 %We would like to emphasize that the function  $\psi(z)$
 %depends on a single scaling variable $z$.
 %The existence of such a solution is not evident in advance.

%The expression (\ref{eq:r20}) relates the differential
% cross section for the production of  the inclusive particle $m_{1}$
% and the average particle  multiplicity density $\rho_A(s,\eta)$
% with the scaling function $\psi(z)$.
The function is normalized as
\begin{equation}
\int_{z_{min}}^{\infty} \psi(z) dz = 1. \label{eq:b6}
\end{equation}
The equation allows us to give the physical meaning of the
function $\psi$ as a probability density to form a particle  with
the corresponding value of the variable $z$.

 In accordance with the approach suggested in \cite{Z99},
 the variable $z$ is taken in the form
 (\ref{eq:r28})
 as a simple physically meaningful variable reflecting
   self-similarity and fractality as a general
    pattern of hadron production at high energies
\begin{equation}
z = \frac{ \sqrt{ {\hat s}_{\bot} }} {\Omega \cdot \rho_A(s) }.
\label{eq:r28}
\end{equation}
Here $\sqrt{ {\hat s}_{\bot} }  $ is the minimal transverse
energy of  colliding constituents necessary to produce a real
hadron in the reaction (\ref{eq:r1}). The factor $\Omega$ is
given by (\ref{eq:r8})
 and   $\rho_A(s) =\rho_A(s, \eta)|_{\eta=0}$.
%The transverse energy consists of two parts which represent the
%transverse energy of the inclusive particle and its recoil.
 The form of $z$ determines its variation range $(0,\infty) $.
 These values are scale independent and
 kinematically accessible at any energy.

  One of the features of the procedure to construct $\psi(z)$   and $z$
described above  is the joint use of the experimental quantities
characterizing hard ($Ed^3\sigma/{dq^{3}}$) and soft ($\rho_A(s,
\eta)$) processes of particle interactions.
 % Therefore, there is a real problem
 % for a theoretical description of $z$-scaling in the
 % framework of perturbative QCD. We would like to note that
 % $z$-construction is not direct mathematical consequence of
 % parton model of strong interaction but it is a new
 % self-similarity pattern motivated by parton-parton and string-like
 % scenarios of particle interactions.

   Let us clarify the physical meaning of
   the variable  $z=\sqrt{\hat s_{\bot}}/(\Omega\rho_A)$.
   The value $ \sqrt {\hat s_{\bot}}$ is the minimal transverse energy of
  colliding constituents necessary to produce a real hadron in the reaction
  (\ref{eq:r1}). It is assumed that two point-like and massless
  elementary constituents interact with each other in the initial state and
  convert into real hadrons in the f{i}nal state. The conversion is not
  instant process and is called  hadronization or particle formation.
  The microscopic space-time picture of the hadronization is not
  understood enough at present time.
  We assume that number of hadrons produced
  in the hard interaction of constituents
  is proportional to $\rho_A$. Therefore the value
  $ \sqrt {\hat s_{\bot}}/\rho_A$ corresponds to the energy density
  per one hadron produced in the sub-process.
  The factor $\Omega $ is relative number of all initial
  configurations containing the constituents which carry the
  momentum fractions $x_1$ and $x_2$. This factor thus represents
  a tension in the considered sub-system with respect to the
  whole system.
  Taking into account the qualitative scenario of hadron formation
  as a conversion of a point-like constituent into a real hadron
  we interpreted the variable $z$ as particle formation length.

{\section{PROPERTIES of $z$-SCALING}}

In this section we discuss properties of the $z$-scaling for direct photons produced
 in $pp$, $\bar pp$ and $pA$
collisions. They are the energy  and angular independencies of data $z$-presentation, the power law of the scaling
function at very high-$p_T$ and  $A$- dependence of $z$-scaling. All properties are asymptotic ones because they
reveal themselves at extreme conditions (high $\sqrt s$ and $p_T$). Numerous experimental data obtained at ISR
\cite{WA70}-\cite{R807}, SpS \cite{UA6p}-\cite{UA2}
%\cite{UA6p,UA6bp,UA1,UA2}
and Tevatron \cite{E704}-\cite{E706g}
%\cite{E704,CDF1pho,CDFpho,D0pho,E706g}
were used in the analysis.

{\subsection{Energy independence}}

The energy independence of data $z$-presentation means that the
scaling function $\psi(z)$ has the same shape for different
$\sqrt s$ over a wide $p_T$ range.

Figures 1(a,c)  show the dependence  of the cross section of
direct photon production in $pp$ and $\bar pp$ interactions on
transverse momentum $p_T$ at different $\sqrt s $ over a central
rapidity range. We would like to note that the data cover a wide
transverse momentum range, $p_T = 19-63~GeV/c$ and
$p_T=24-1800~GeV/c$ for $pp$ and $\bar pp$, respectively.

\begin{figure}
\begin{center}
\hspace*{-8cm}
\includegraphics[width=6.5cm]{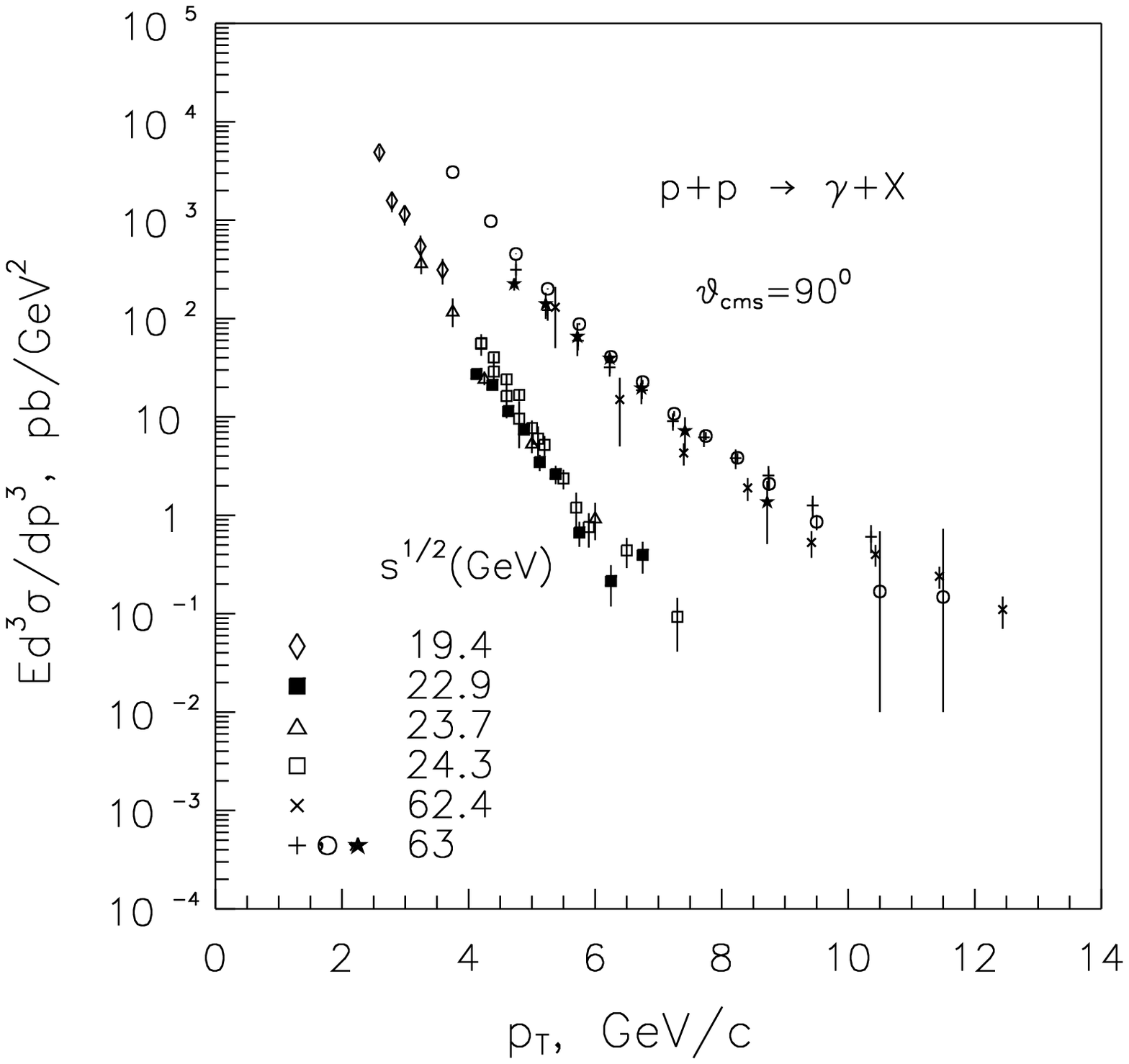}
\vskip -6.cm \hspace*{7cm}
\includegraphics[width=6.5cm]{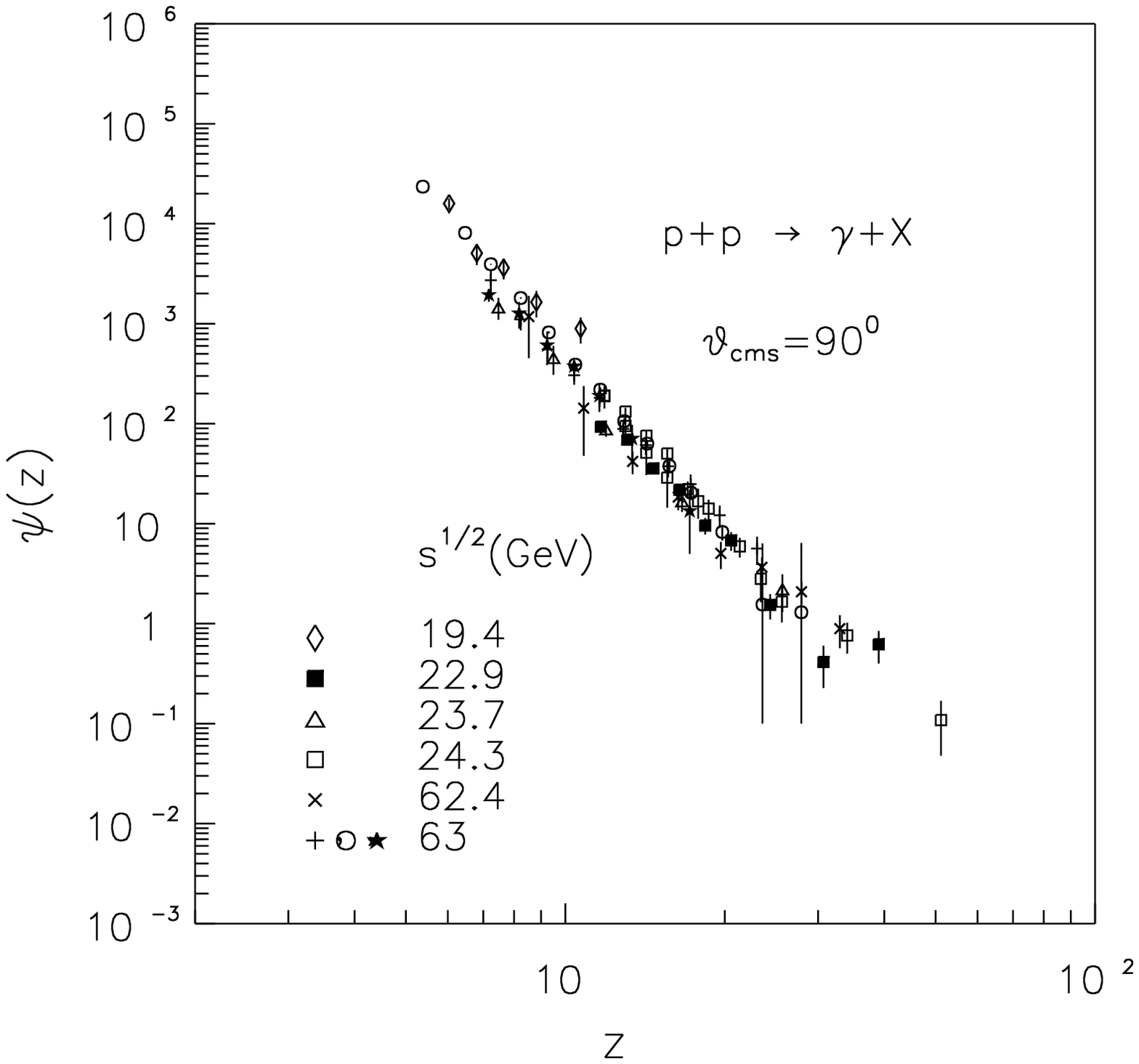}

%\caption{The CERN $\bar{\rm p}$ complex.}
\vskip 0.5cm

\hspace*{1cm} a) \hspace*{7cm} b)
\end{center}
%\end{figure}

%\begin{figure}
\begin{center}
\hspace*{-8cm}
\includegraphics[width=6.5cm]{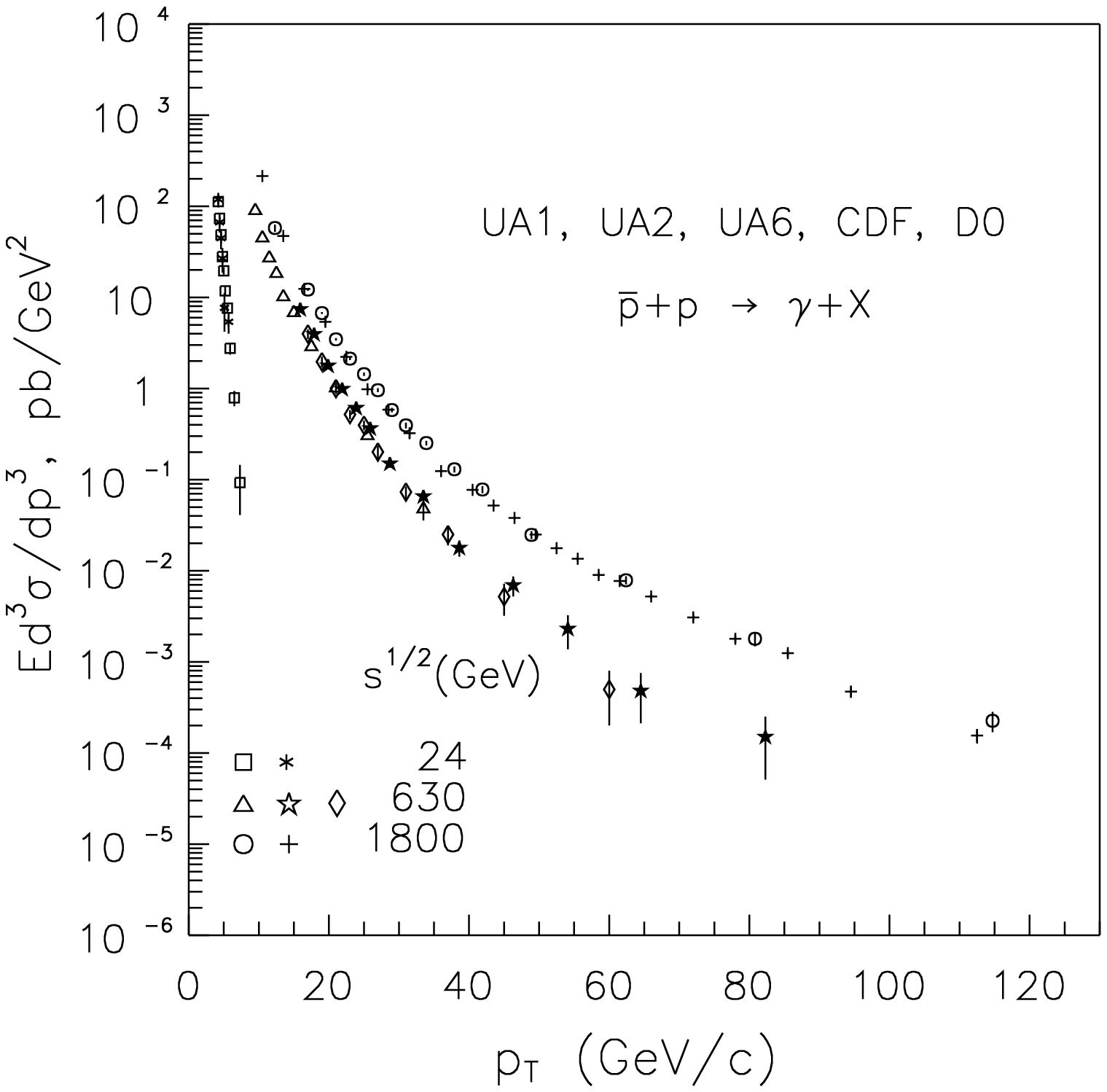}
\vskip -6.5cm \hspace*{7cm}
\includegraphics[width=6.5cm]{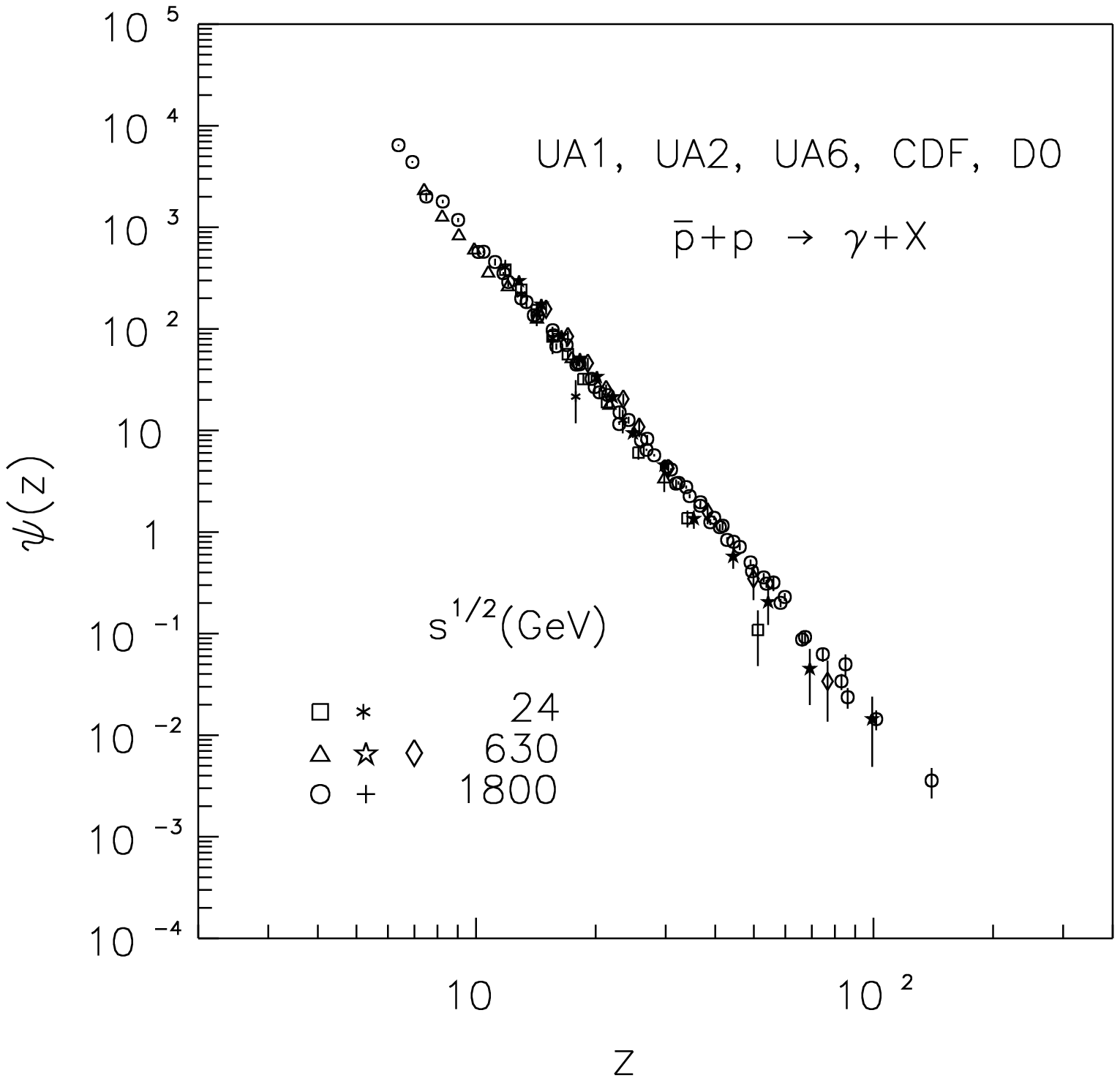}

\hspace*{1cm} c) \hspace*{7cm} d)
\caption{
 Dependence of inclusive cross sections of direct photon
production on transverse momentum in $pp$ (a)  and $\bar pp$ (c)
collisions at $\sqrt s = 19-1800~GeV$. Experimental data are
taken from
%\cite{WA70,R108,R110,NA24,R806,R807,UA6p,UA6bp,UA1,UA2,E704,CDF1pho,CDFpho,
%D0pho,E706g}.
\cite{WA70}-\cite{E706g}. The corresponding scaling functions
$\psi(z)$ for $pp$  and $\bar pp$  collisions are shown in (b) and
(d).}
\end{center}
\end{figure}

Let us note some features of the photon  spectra. The first one
is the  strong  dependence of the cross section on energy $\sqrt
s$. The second feature is a tendency
 that  the difference between photon yields increases
 with the transverse momentum $p_T$ and the energy $\sqrt s$.
 The third one is a non-exponential behavior of the spectra
 at $q_{T}>4~GeV/c$.

 Figures 1(b,d) show $z$-presentation of the same data sets.
 Taking into account the experimental errors we can conclude that
 the scaling function $\psi(z)$ demonstrates energy
 independence over a wide energy and transverse momentum
 range at $\theta_{cms} \simeq 90^0$.

{\subsection{Angular independence}}

The angular independence of data $z$-presentation means that the scaling function $\psi(z)$ has the same shape for
different values of the  angle $\theta_{cms} $ of produced photon over a wide $p_T$ range and $\sqrt s$. Taking
into account the energy independence of $\psi(z)$ it will be enough to verify the property at some $\sqrt s$.

To analyze the angular dependence of the scaling function
$\psi(z)$ we use some data sets. The first one obtained at
Tevatron \cite{D0pho} includes the results of measurements of the
invariant cross section $Ed^3\sigma/dq^3$  at  $\sqrt s =
1800~GeV$ over a momentum and angular ranges of
$p_T=10-115~GeV/c$ and $0.0<|\eta|<2.5$. The second one is the
E706 data set \cite{E706g} for direct photons produced in $pp$
collisions at $\sqrt s = 31.6$ and $38.8~GeV$ and in the rapidity
range $(-1.0,0.75)$.

\begin{figure}{}
\begin{center}
\hspace*{-8cm}
\includegraphics[width=6.5cm]{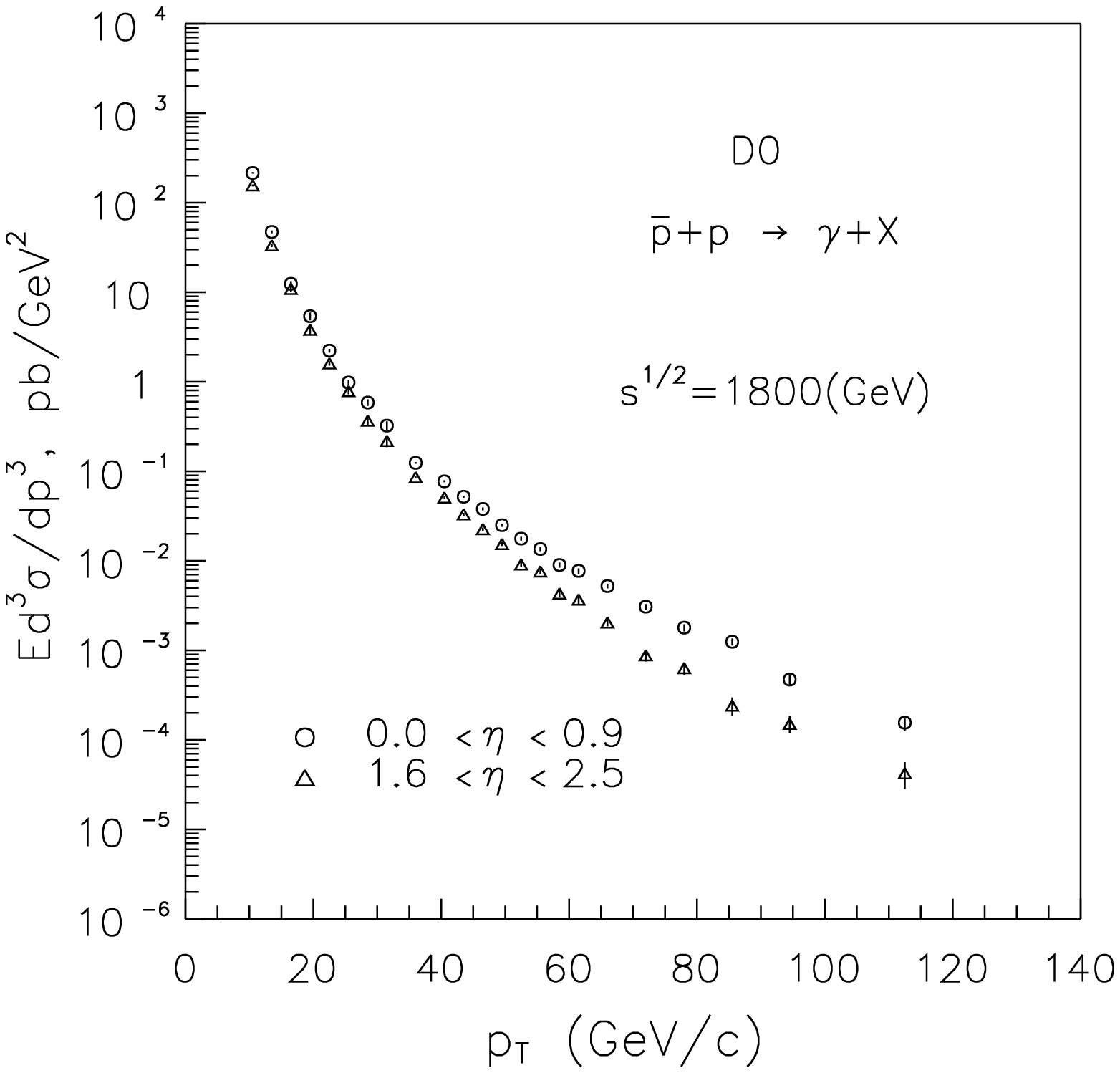}
\vskip -6.5cm \hspace*{7cm}
\includegraphics[width=6.5cm]{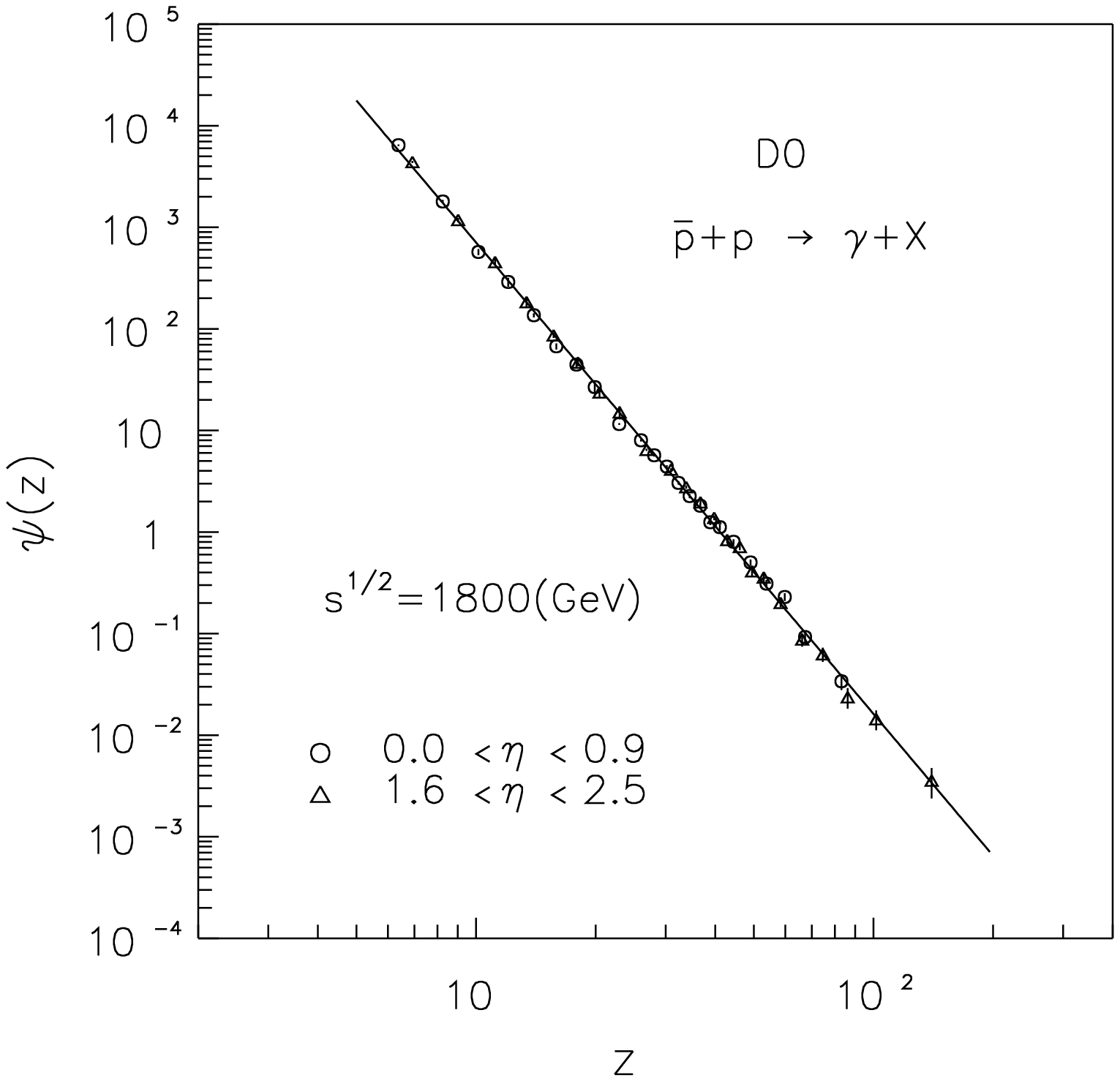}
%\vskip 0.5cm

\hspace*{1cm} a) \hspace*{7cm} b) \caption{
 (a) Dependence of
the inclusive cross section of direct photon production on transverse
momentum in $\bar pp$ collisions for  pseudorapidity intervals (0.0,0.9)
 and (1.6,2.5) at $\sqrt s = 1800~GeV$. The experimental data on cross sections
obtained by D0 Collaboration \cite{D0pho} are used. (b) The corresponding
scaling function $\psi(z)$. }

\end{center}

%\end{figure}

%\begin{figure}
\begin{center}
\hspace*{-8cm}
\includegraphics[width=6.5cm]{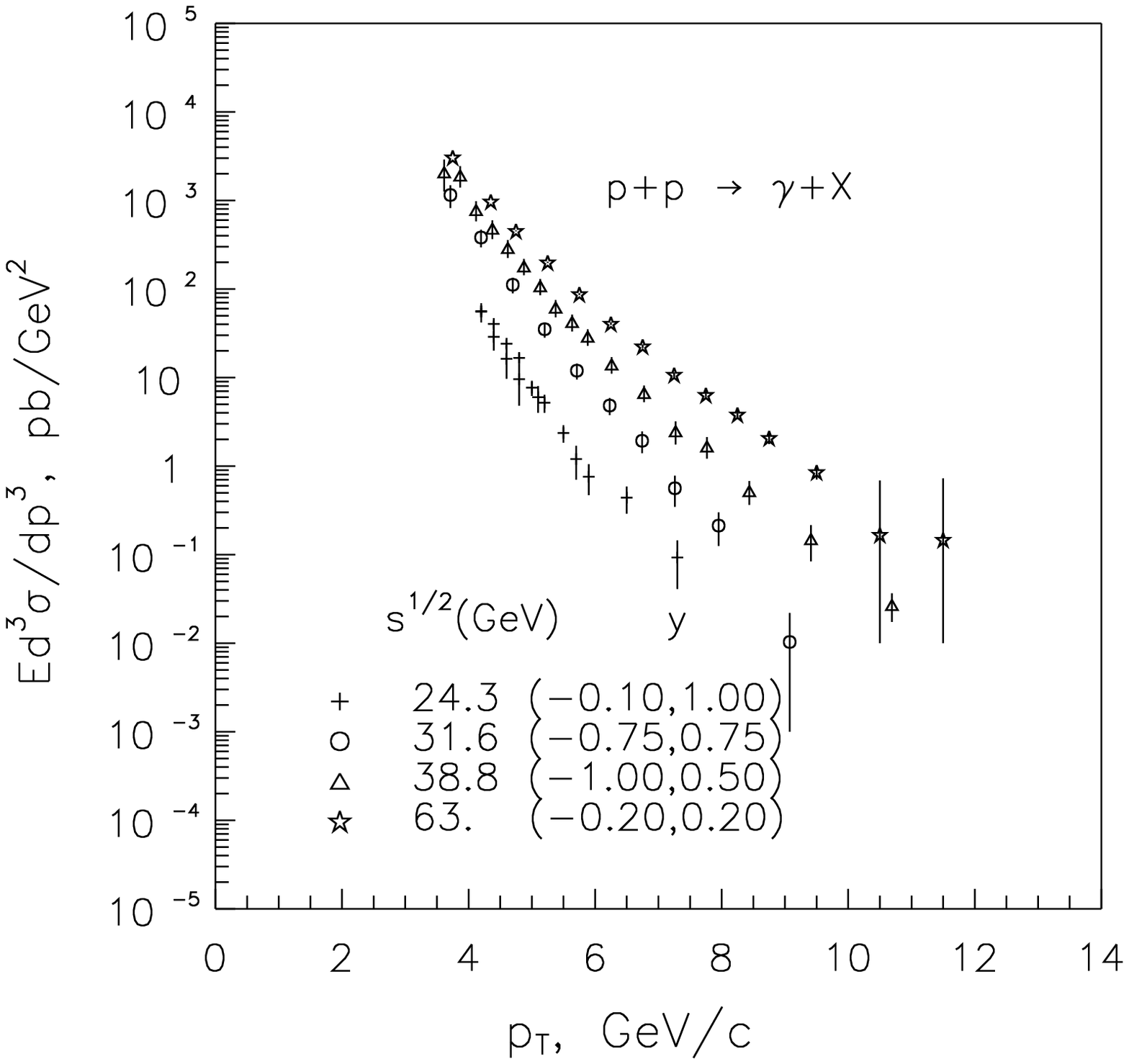}
\vskip -6.cm \hspace*{7cm}
\includegraphics[width=6.5cm]{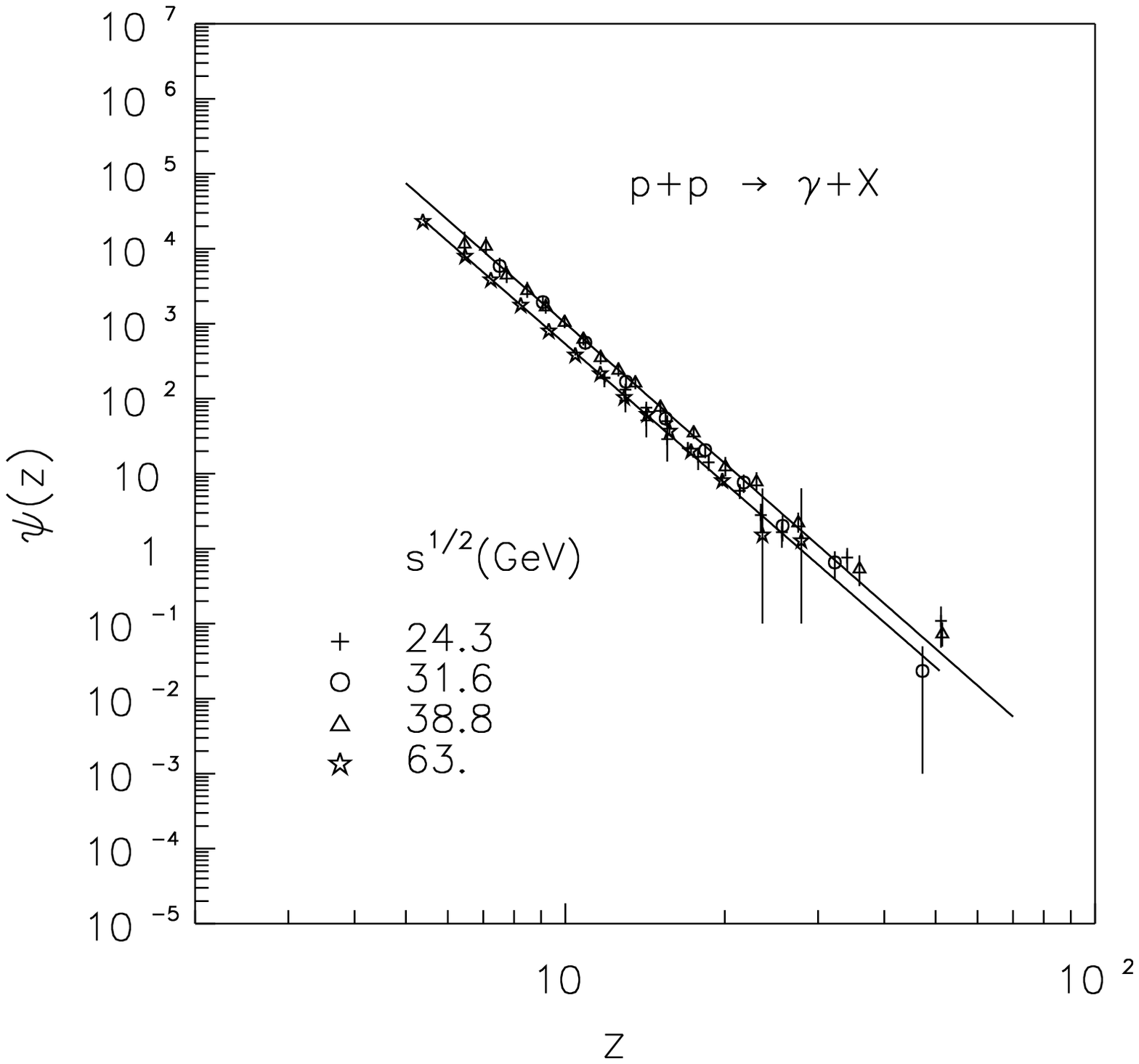}

\hspace*{1cm} c) \hspace*{7cm} d) \caption{
 (a) Dependence of inclusive cross sections of direct photon
production on transverse momentum in $pp$ collisions at $\sqrt s
= 24.3,31.6, 38.8$ and $63.~GeV$. Experimental data $+, \star $
and $\circ, \triangle $ are taken from \cite{R806,UA6p} and
\cite{E706g}, respectively. (b) The corresponding scaling
function $\psi(z)$.}
\end{center}
\end{figure}

Figures 2(a) and 3(a) show the dependence of the cross section of
direct-$\gamma$ production in $\bar pp$  and $pp$ collisions on
transverse momentum at the fixed $\sqrt s$ and for different rapidity
intervals. A strong angular dependence of the cross section was
experimentally observed for D0 data and was found to be much
smaller for E706 data. The last data have been averaged over
the central rapidity range and therefore the angular dependence of
the data is weak.

Figures 2(b) and 3(b) demonstrate $z$-presentation of the same
data sets. The obtained results show that the function $\psi(z)$
is independent of the angle $\theta_{cms}$ over a wide range of
transverse momentum  $p_T$ and energy $\sqrt s$. This is the
experimental confirmation of the angular scaling of data
$z$-presentation.

We would like to note that absolute normalization factors for
\cite{WA70,R806} and \cite{E706g} data sets are found to be
different. The ratio is about factor 0.5.

{\subsection{Power law}}

Here, we discuss a new feature of data $z$-presentation for
direct-$\gamma$ production. This is the power law of the scaling
function, $\psi(z) \sim z^{-\beta}$.

As seen from  Figures 2(b) and 3(b) the data sets demonstrate a
linear $z$-dependence of $\psi(z)$ on the log-log scale at high
$z$. The quantity  $\beta $ is a slope parameter.

Taking into account the accuracy of the available experimental
data, we can conclude that the behavior of $\psi(z)$ for direct
photons  produced in $\bar pp$ and $pp$
 collisions reveals a power dependence and the value of the
slope parameter is independent of the energy $\sqrt s$  over a
wide range of high transverse momentum.  It was also found that
 $\beta_{pp}^{\gamma} > \beta_{\bar pp}^{\gamma}$.

% The mean values of
% $\beta_{pp}^{\gamma}$ and $\beta_{\bar pp}^{\gamma}$
%are found to be 5.91 and  5.48, respectively.

Direct photons are mainly produced in $p-p$  and $\bar p-p$
collisions through the Compton  and annihilation processes,
respectively. This fact causes  different
values of the slope parameters $\beta_{pp}^{\gamma}$ and
$\beta_{\bar pp}^{\gamma}$.

% Thus we can conclude based on the obtained results
% that  behavior of $\psi(z)$
% for $\pi^0$-meson, direct-$\gamma $  and  $jet$ production
% at high $z$
% reveals the power dependence with high accuracy.
%
%   The value of the slope
%   parameter is independent of the colliding energy
%   $\sqrt s$ and  the angle of produced particles  over a wide
%   range of high transverse momentum.

 The existence of the power law, $\psi(z) \sim z^{-\beta }$,
 means, from our point of view, that
 the mechanism of particle formation reveals fractal behavior.

{\subsection{A-dependence}}

A study of  $A$-dependence of particle production in $hA$ and
$AA$ collisions is traditionally connected with  nuclear matter
influence  on particle  formation. The difference  between  the
cross sections of particle production on free and bound nucleons
is normally considered as an indication of unusual physics
phenomena like EMC-effect
%\cite{EMC,SLAC},
$J/\psi$-suppression
%\cite{NA38,NA50},
and Cronin effect \cite{Cronin}.
%enhanced $A$-dependence of charged hadron production in $p-A$ collisions \cite{Cronin}).

A-dependence of $z$-scaling for particle production in $pA$
collisions was studied in \cite{Z01}. It was established
$z$-scaling for different nuclei ($A=D-Pb$) and type of produced
particles ($\pi^{\pm,0}, K^{\pm}, \bar p$). The symmetry
transformation of the scaling function $\psi(z)$ and variable $z$
under the scale transformation $z\rightarrow \alpha_A z$, $\psi
\rightarrow \alpha_A^{-1} \psi $ was suggested to compare the
scaling functions for different nuclei. It was found that $\alpha$
depends on the atomic number only and can be parameterized by the
formula $\alpha(A)=0.9A^{0.15}$ \cite{Z01}.

We use the parameterization  $\alpha (A)$  to study $A$-dependence
of direct photon production in $pA$ collisions. New data
\cite{E706g} obtained by E706 Collaboration are used in the
analysis. The experimental cross sections have been measured for
$pBe$ and $pCu$  collisions at $\sqrt s = 31.6$ and $38.8~GeV$
and cover the $p_T$-range $(3-11)~GeV/c$.

Figure 4(a) demonstrates the spectra of photons produced in
proton-nucleus collisions. As seen from Figure 4(a) the
$p_T$-spectra shows the strong energy dependence. The difference
between spectra at $\sqrt s = 31.6$ and $38.8~GeV$ increases with
$p_T$. The $z$-presentation of the same data is shown in Figure
4(b). The scaling functions for both targets, $Be$ and $Cu$,
coincide each other.  This is  the direct confirmation that a nuclear
effect for direct photon production can be described by the same
function $\alpha (A)$ as for hadrons produced in proton-nucleus
collisions \cite{Z01}. The shape of the scaling functions is
found to be a linear one on the log-log scale for both cases. The
fit of the data is shown by the solid line in Figure 4(b).

\begin{figure}
\begin{center}
\hspace*{-8cm}
\includegraphics[width=6.5cm]{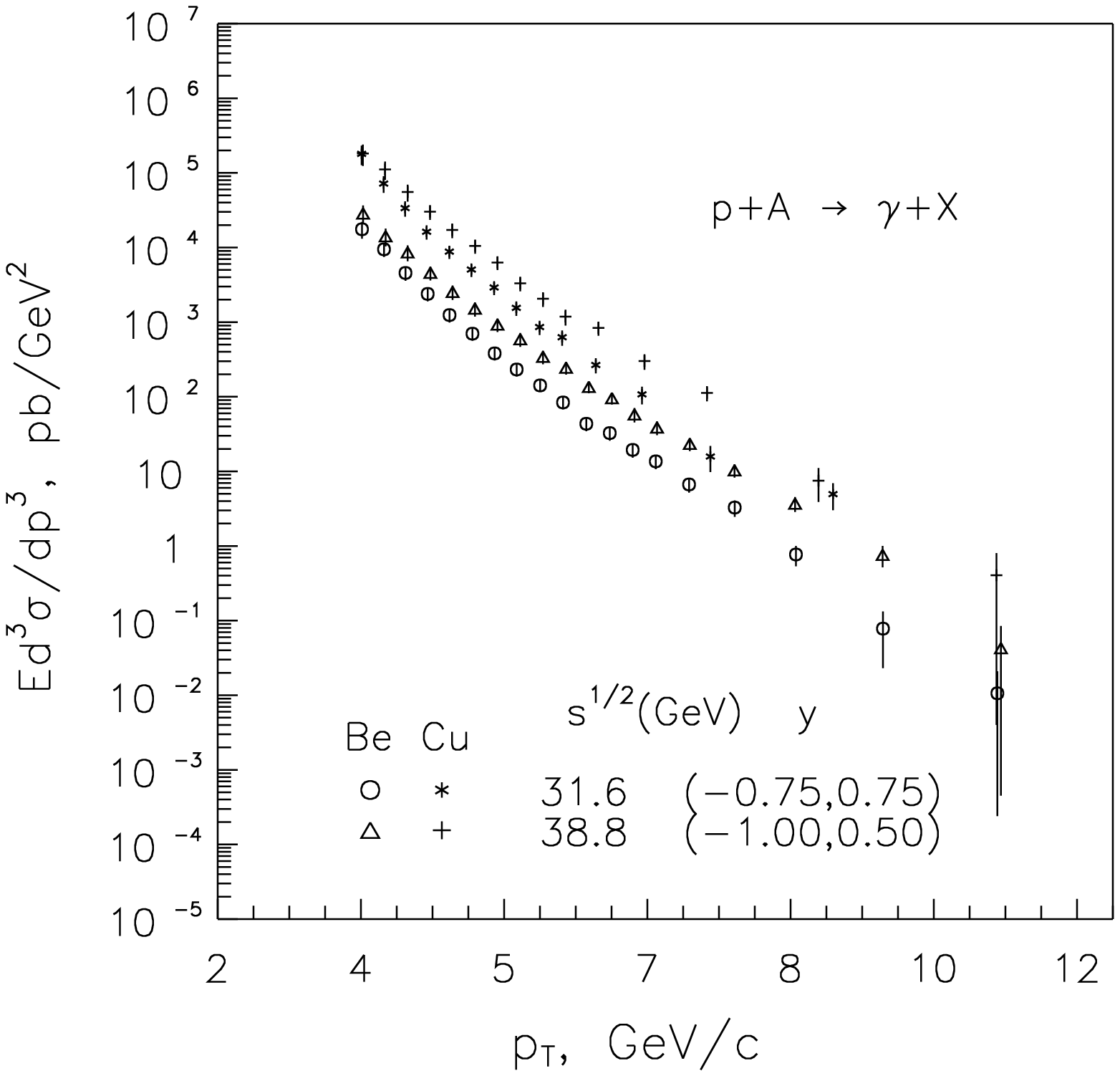}
\vskip -6.cm \hspace*{7cm}
\includegraphics[width=6.5cm]{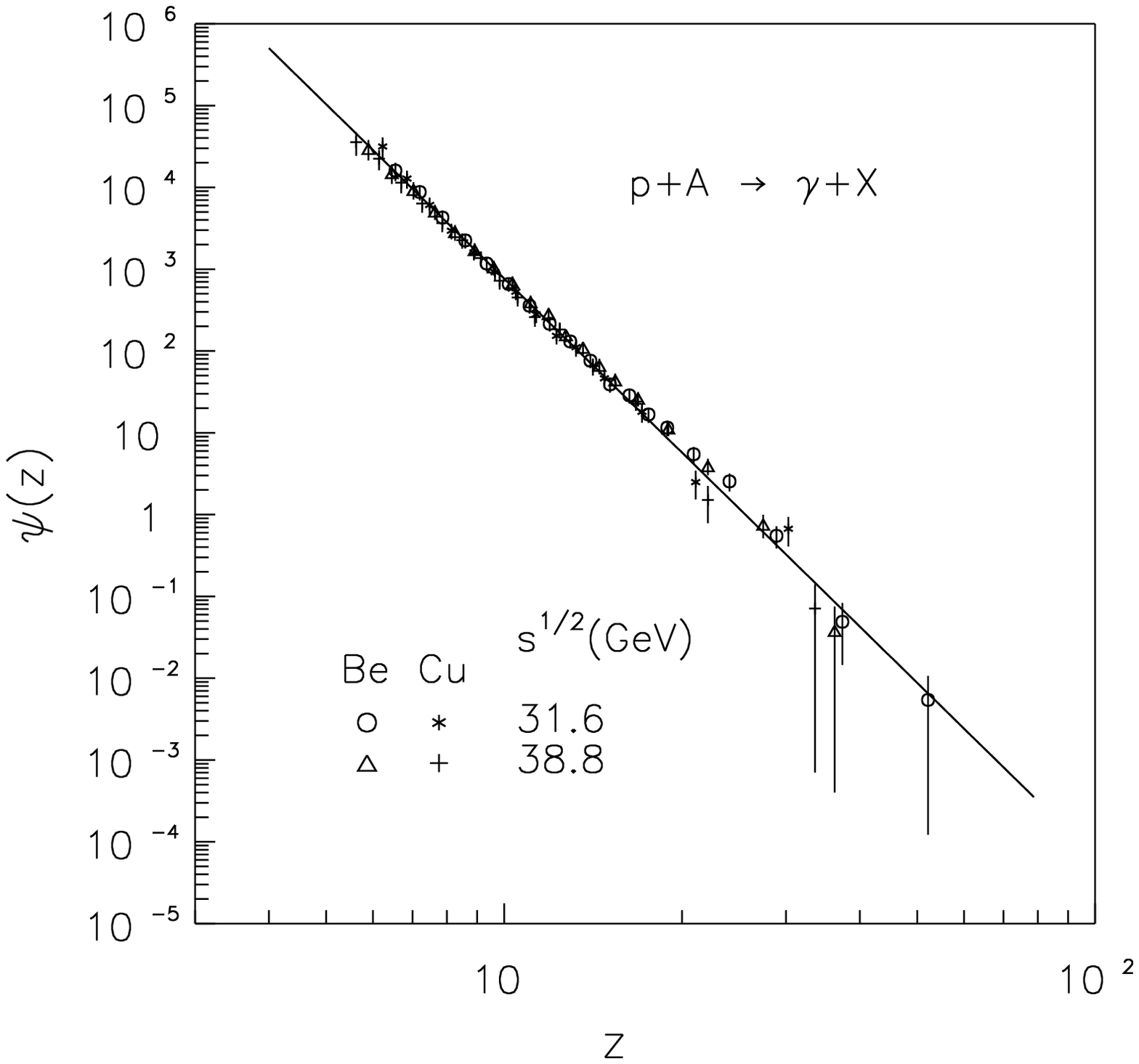}

%\caption{The CERN $\bar{\rm p}$ complex.}
\vskip 0.5cm

\hspace*{1cm} a) \hspace*{7cm} b)
\end{center}
%\end{figure}

%\begin{figure}
\begin{center}
\hspace*{-8cm}
\includegraphics[width=6.5cm]{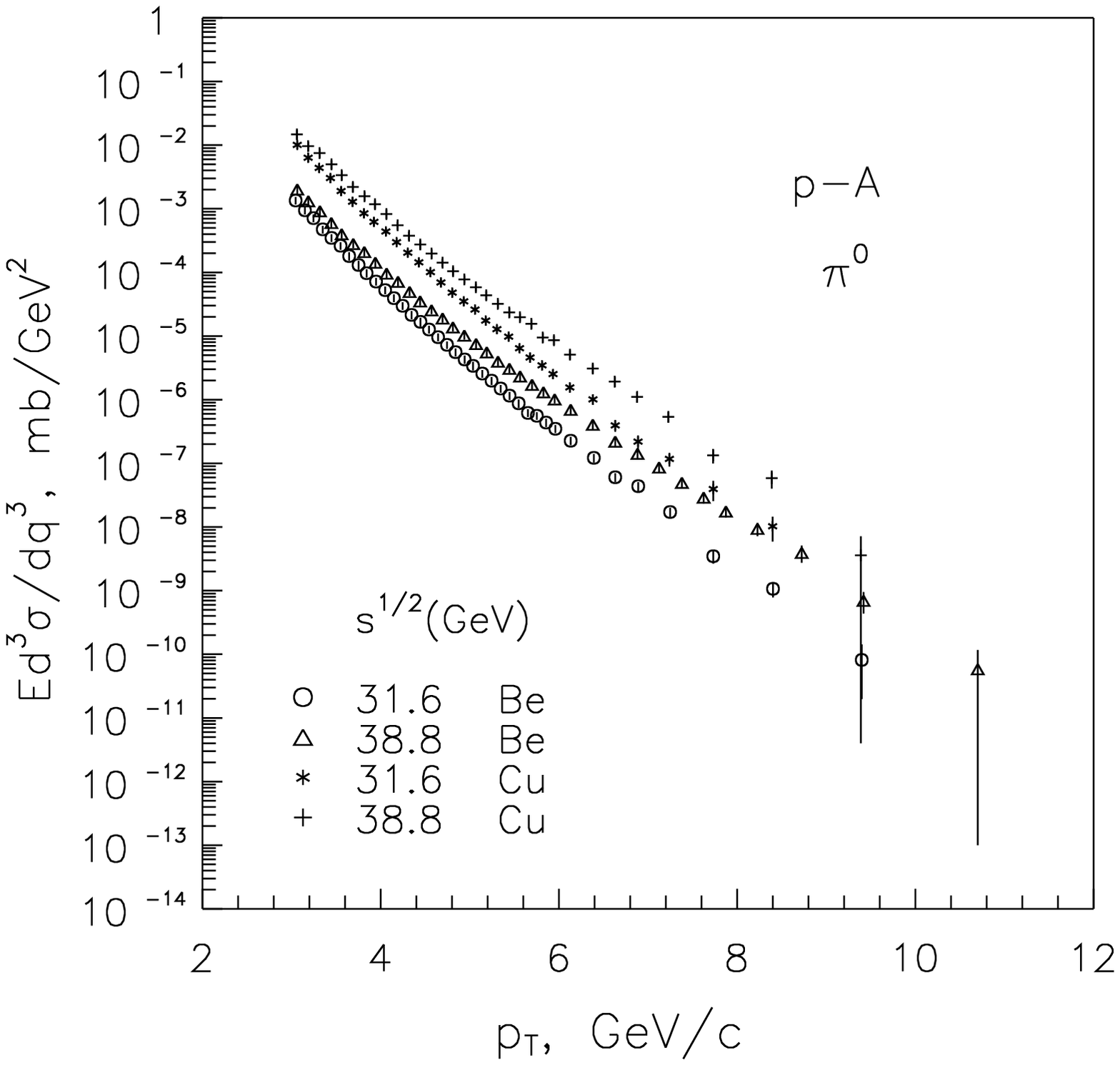}
\vskip -6.5cm \hspace*{7cm}
\includegraphics[width=6.5cm]{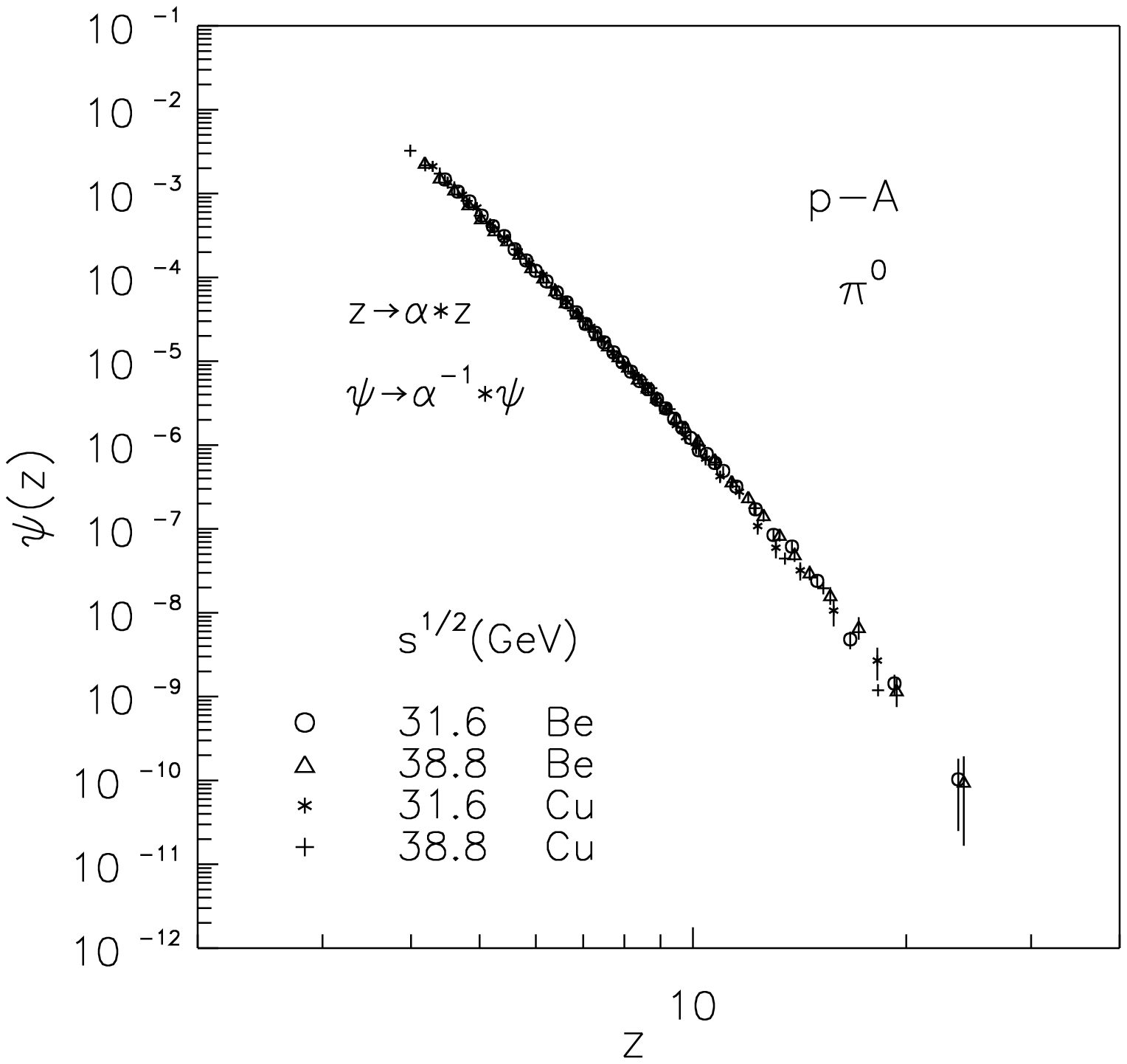}

\hspace*{1cm} c) \hspace*{7cm} d) \caption{
 Dependence of inclusive cross sections of direct photon (a) and
$\pi^0$-meson (c)  production on transverse momentum in $pBe$ and
$pCu$ collisions at $\sqrt s = 31.6$ and $38.8~GeV$. Experimental
data are taken from \cite{E706g}. The corresponding scaling
functions $\psi(z)$ for $\gamma$ and $\pi^0$ are shown in (b) and
(d). }
\end{center}
\end{figure}

The value of the slope parameter $\beta_{pBe}^{\gamma}$ is
constant over a wide $p_T$ range and equal to  7.07. The fact
means that the nuclear matter changes the probability of photon
formation with different formation length  $z$ and does not
change  the fractal dimension of the mechanism of photon
formation (photon "dressing").

Taking into account an experimental accuracy of data used in the
analysis, the obtained results show that the fractal dimension
$\delta$ and  the slope parameter $\beta$ is independent of A.
Therefore the experimental investigations of $A$-dependence of
$z$-scaling for direct photons  produced in hadron-nucleus
collisions at RHIC and LHC energies are very important to obtain
any indications on nuclear phase transition and  formation of QGP.

The main source of the background  for direct photon production
are $\pi^0$ and $\eta^0$-mesons decay. Therefore it is important
to obtain a reliable estimation of the background. This can be
done by using the scaling function of $\pi^0$-mesons for
calculation of cross sections at the corresponding  energy. We use
experimental data \cite{E706g} on $\pi^0$-meson cross sections to
construct the scaling function $\psi(z)$.

Figure 4 shows the  dependence of inclusive cross section of
$\pi^0$-meson produced in proton-nucleus collisions on transverse
momentum and the results of $z$-presentation of the same data
sets. The values of the fractal dimension $\delta$ were found to
be different, 0.5 and 0.8,  for $\pi^0$ and $\gamma$  production,
respectively. The values of the slope parameter $\beta$ of
$\psi(z) \sim z^{-\beta}$ were found to be different as well.

%New data on high-$p_T$ cross sections of $\pi^0$-mesons produced in $Au-Au$ collisions at RHIC are presented
%by PHENIX Collaboration  in \cite{Gabor}. The  pion spectra for different centralities are shown in Figure 18
%(a). We compared the scaling function  for minimum bias  data set with  function  found at ISR and SpS
%energies. The obtained results are shown in Figure 18(b).
%Note that the dramatic change of the value of the
%fractal dimension $\delta $ was found. The scaling function at ISR and SpS energies corresponds to
%$\delta=0.5$ whereas  at RHIC energies the value of $\delta$ is found to be 3.5. This is indication that
%nuclear environment changes essentially the mechanism of particle formation. The increase of fractal dimension
%$\delta$ means that mechanism of  multiple scattering play an important role of particle formation in nuclear
%medium created at RHIC energies. We suggest to use the change of the fractal dimension ("$\delta$-jump") as
%signature of nuclear matter transition. Therefore it is of interest to investigate the energy dependence of
%fractal dimension $\delta (s)$ and determine the shape of the dependence.

{\section{DIRECT  $\gamma$, $\pi^0$ and $\eta^0$ PRODUCTION}}

The properties of $z$-scaling found for direct-$\gamma$, $\pi^0$-
and $\eta^0$-meson production allow us to calculate spectra of
photons produced in $pp$ and $pA$ collisions at RHIC and LHC
energies.

{\subsection{$pp$ and $\bar p$ collisions }}

Results of our analysis of numerous experimental data on direct photon production
in $pp$  and $\bar pp$
collisions  in the framework of $z$-scaling scheme show that the fractal dimension
$\delta$ is independent of
energy $\sqrt s$ over a wide range of transverse photon momentum. Therefore the violation of $z$-scaling could
give indications on modification of the mechanism of direct photon formation by a new type of interaction beyond
Standard Model. A change of the fractal dimension $\delta$ is suggested to be the quantitative measure of the
$z$-scaling violation.

 Figure 5(a) shows that the scaling functions for direct photon,
 $\pi^0$ and $\eta^0$  reveal the power law $\psi(z) \sim z^{-\beta}$ in high-$z$ range.
 It was found  that $\beta_{pp}^{\pi^0} \simeq \beta_{pp}^{\eta^0}$
 and $\beta_{pp}^{\gamma} >  \beta_{pp}^{\pi^0}$.
 Figure 5(b) demonstrates the power law for direct photon
 production in $\bar pp$ collisions as well. The slope parameter
 $\beta_{\bar pp}^{\gamma}$ is found to be 4.58  and
 $\beta_{pp}^{\gamma} > \beta_{\bar pp}^{\gamma}$.
 As seen from Figure 5(c)  the asymptotic shapes of $\psi(z)$ for $\pi^0$-meson
 production  in $pp$  and $\bar pp$ are different. Both ones have
 a power dependence and $\beta_{pp}^{\pi^0} > \beta_{\bar pp}^{\pi^0}$.
  The properties of the scaling functions for direct
 $\gamma$ and $\pi^0$ were used to estimate the dependence
 of the $\gamma / \pi^0$ ratio of inclusive cross sections on
 transverse momentum at $\sqrt s = 5.5$ and $14.0~TeV$.
 Figure 5(d) shows that the ratio increases with $p_T$ and
 it is different for $pp$ and $\bar pp$ collisions.
 The ratio has the crossover point at $p_T\simeq 60-70~GeV/c$  and   $p_T \simeq
 110-130~GeV/c$ for $pp$ and $\bar pp$ collisions, respectively.

% ************************************
\begin{figure}
\begin{center}
\hspace*{-8cm}
\includegraphics[width=6.5cm]{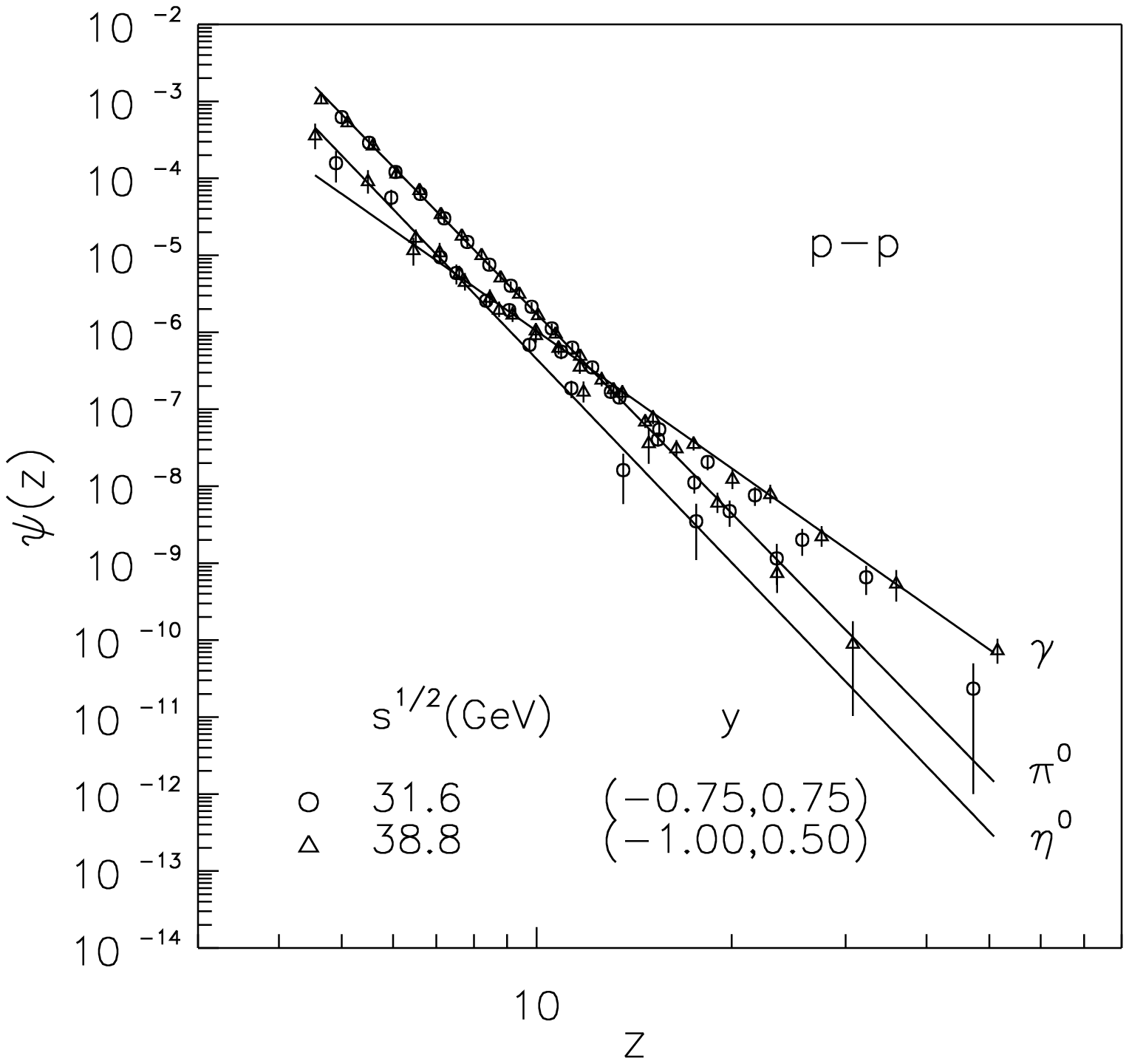}
\vskip -6.cm \hspace*{7cm}
\includegraphics[width=6.5cm]{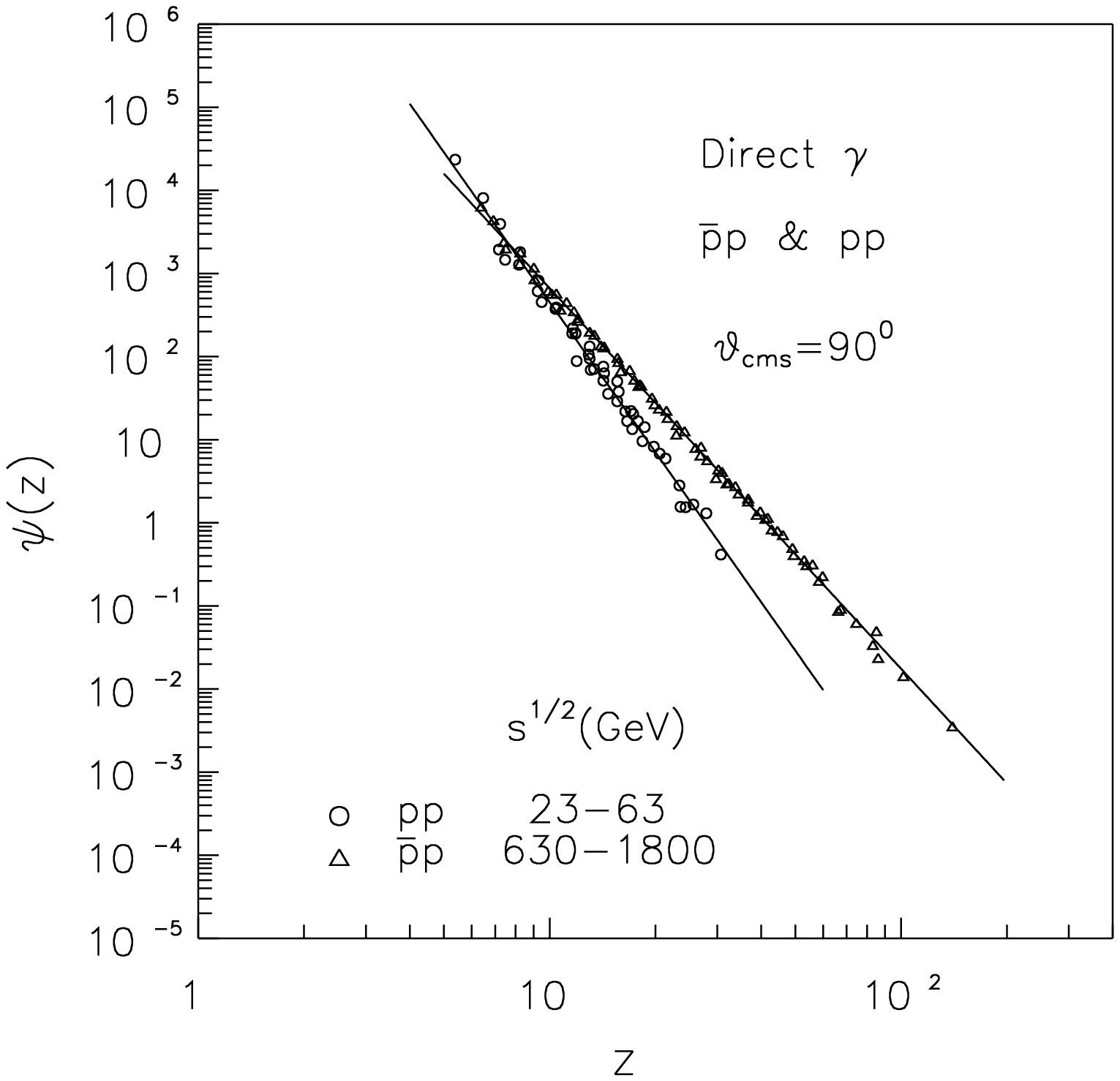}

%\caption{The CERN $\bar{\rm p}$ complex.}
\vskip 0.5cm

\hspace*{1cm} a) \hspace*{7cm} b)
\end{center}
%\end{figure}

%\begin{figure}
\begin{center}
\hspace*{-8cm}
\includegraphics[width=6.5cm]{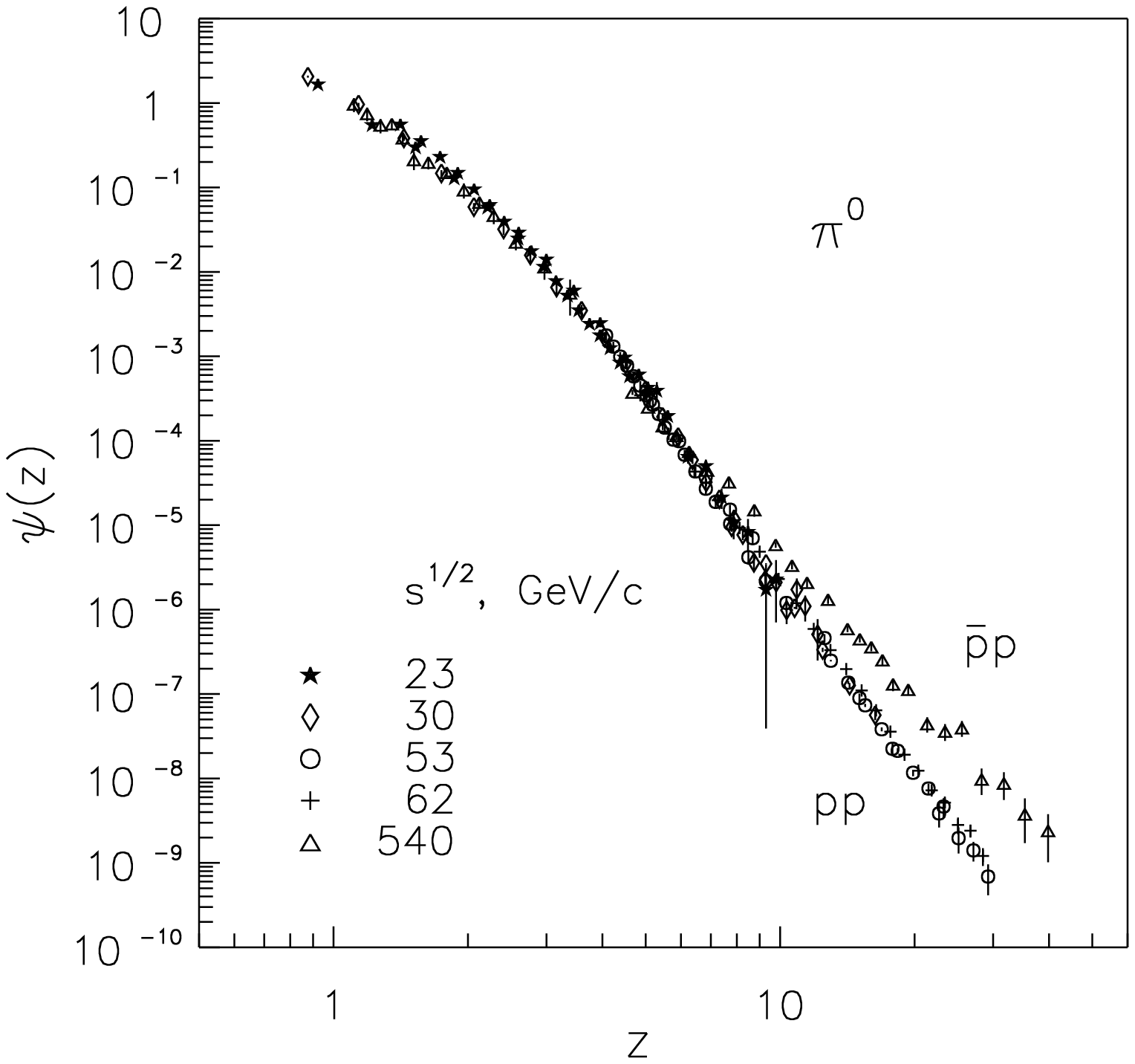}
\vskip -6.cm \hspace*{7cm}
\includegraphics[width=6.5cm]{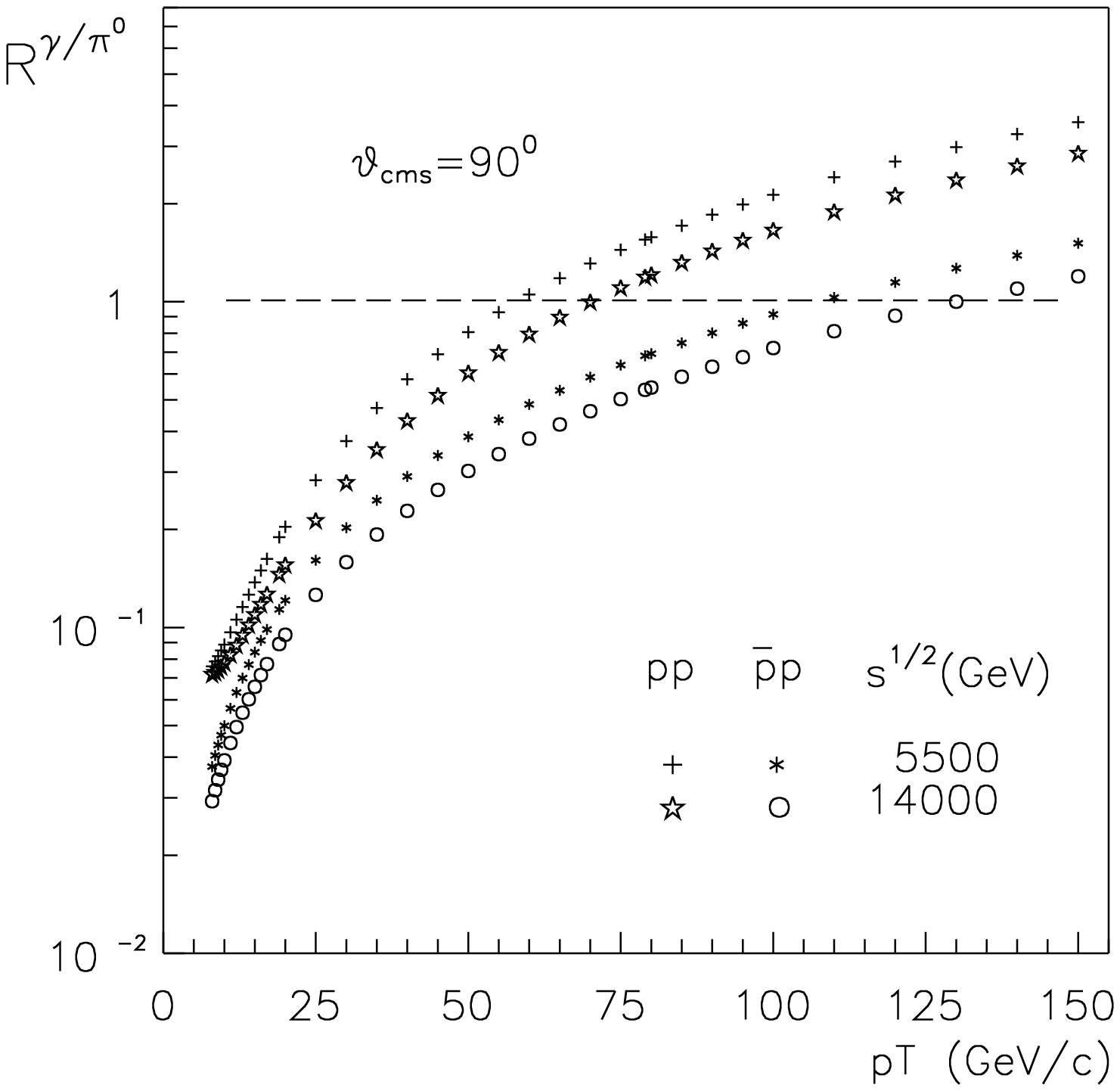}

\hspace*{1cm} c) \hspace*{7cm} d) \caption{ (a) The scaling function of direct photon, $\pi^0$-meson  and
$\eta^0$-meson production in $pp$ collisions. (b) The scaling function  of direct photon production in $pp$
and $\bar pp$ collisions. (c) The scaling function of
 $\pi^0$-mesons  produced  in $pp$ and $\bar pp$ collisions. (d) The $\gamma / \pi^0 $ ratio
 versus transverse momentum in $pp$ and $\bar pp$ collisions at
 $\sqrt s = 5.5$ and $14.0~TeV$.
Experimental data are taken from \cite{WA70}-\cite{Banner}
%\cite{R806,UA6p,E706g,D0pho,CDFpho,Eggert,Angel,Kourk,Lloyd,Banner}
. }
\end{center}
\end{figure}

% **************************************

Figure 6 shows our predictions of the dependence of the inclusive
cross section  $Ed^3\sigma /dq^3$ on transverse momentum  $p_T$
for direct photon (a), $\pi^0$ (b) and $\eta^0$ (c) in $pp$
collisions  at RHIC and LHC energies and at the angle of
$\theta_{cms}=90^0$. The results for the  cross sections at ISR
energy of $\sqrt s = 24-63~GeV$ are also shown for comparison.
The verification of the predictions is very important because it
allows us to confirm or disconfirm the new scaling of photon
production and to determine the region of the scaling validity.

\begin{figure}
\begin{center}
\hspace*{-8cm}
\includegraphics[width=6.5cm]{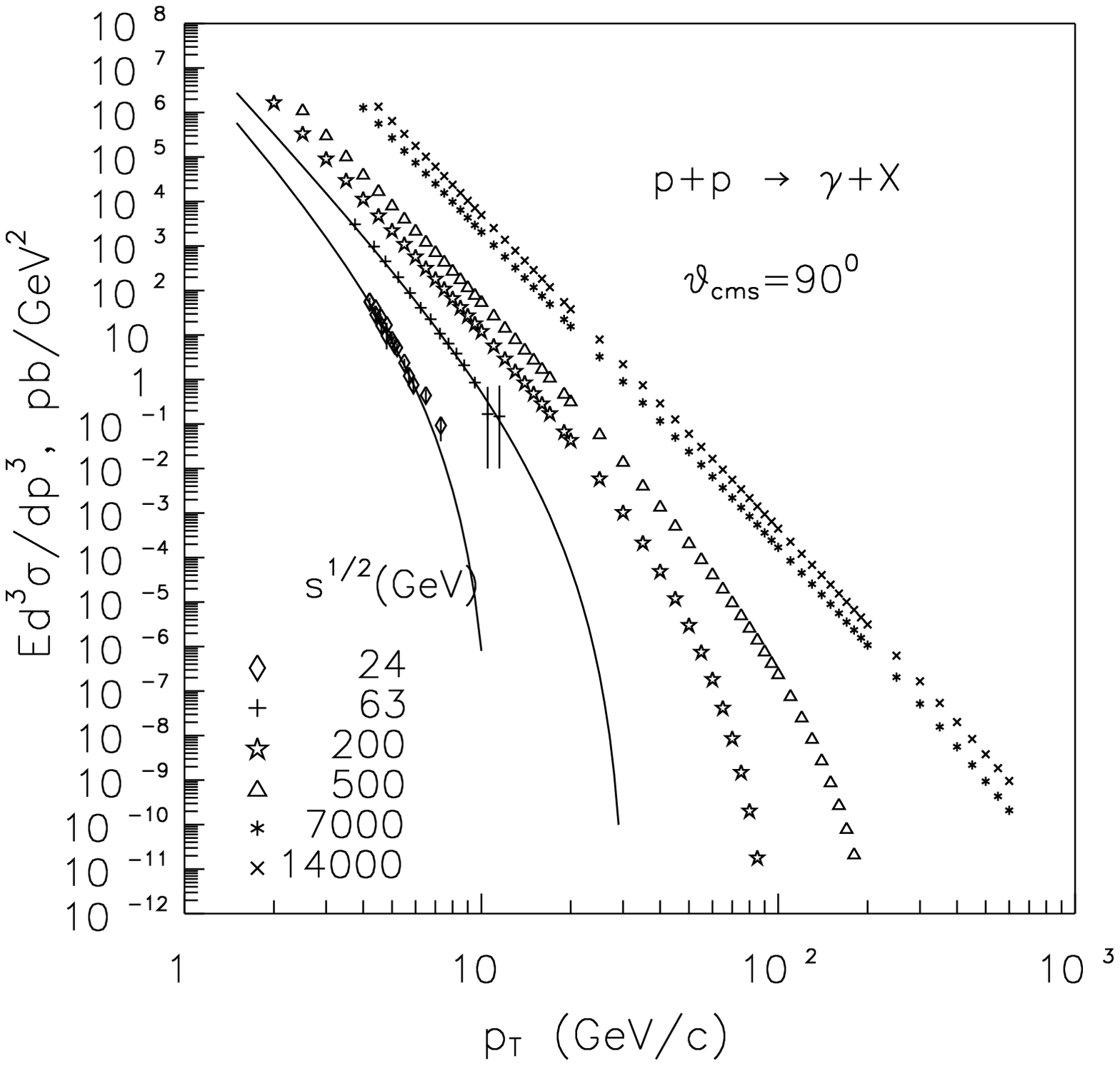}
\vskip -6.cm \hspace*{7cm}
\includegraphics[width=6.5cm]{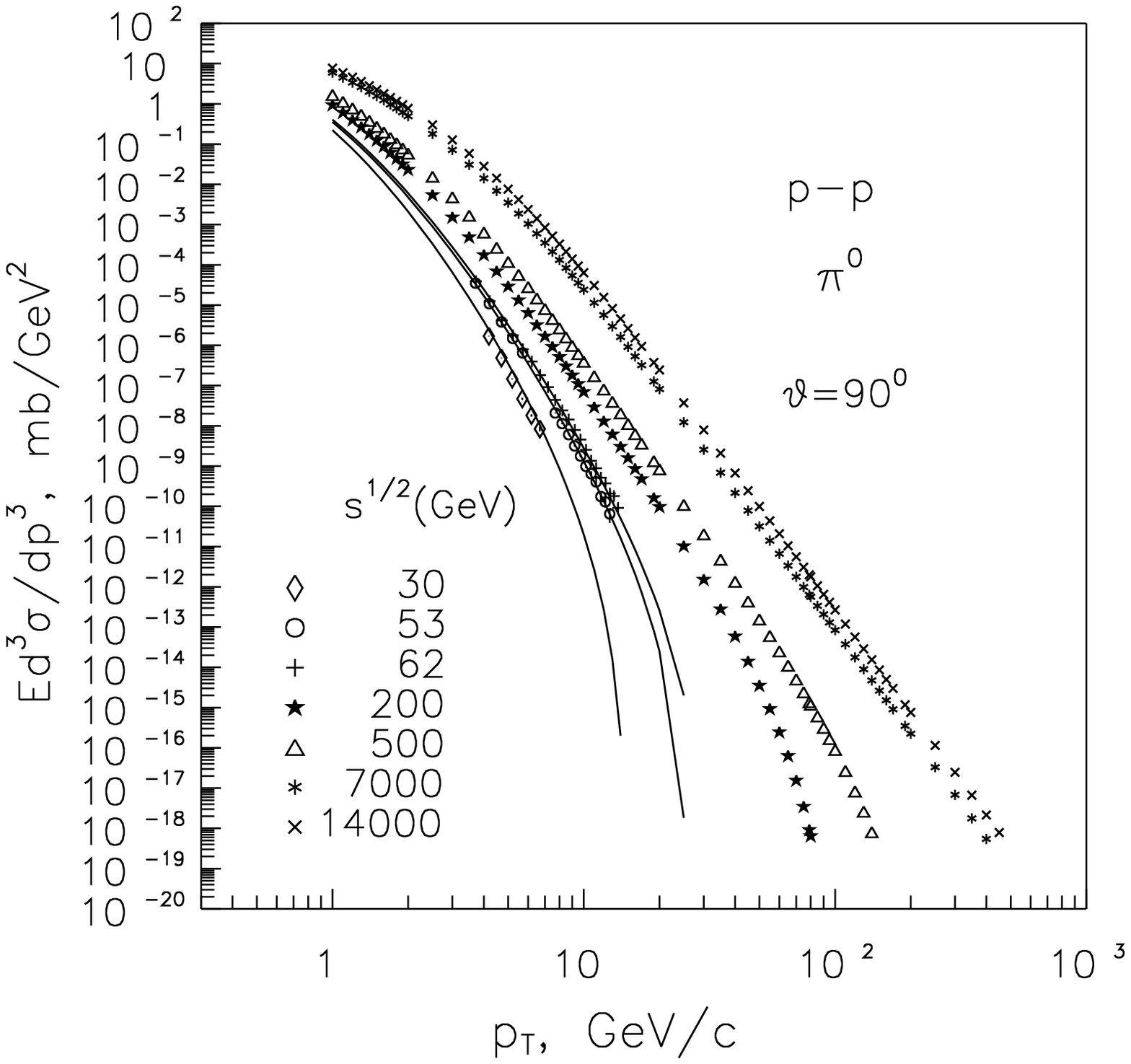}

%\caption{The CERN $\bar{\rm p}$ complex.}
\vskip 0.5cm

\hspace*{1cm} a) \hspace*{7cm} b)
\end{center}
%\end{figure}

%\begin{figure}
\begin{center}
\hspace*{-8cm}
\includegraphics[width=6.5cm]{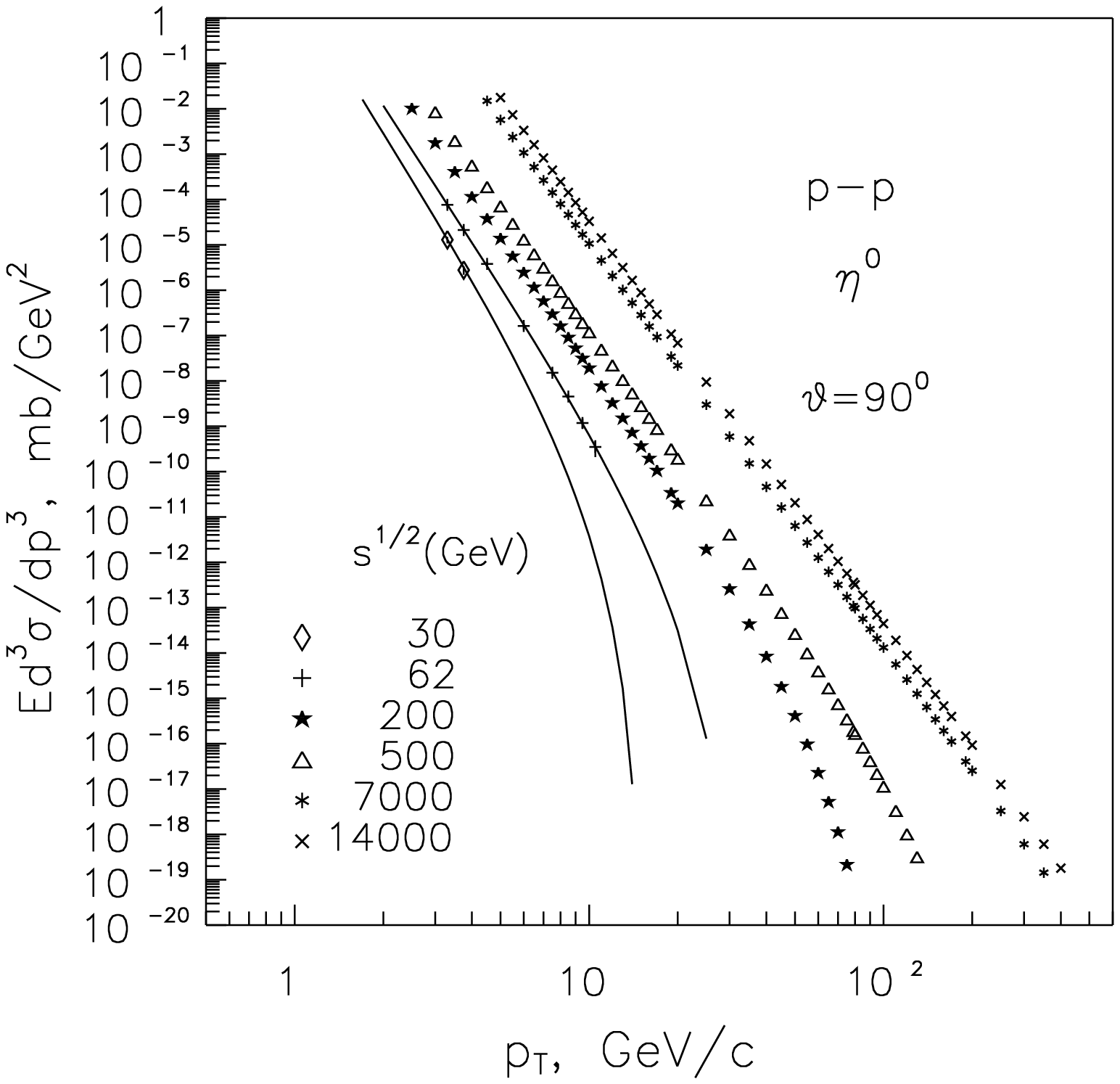}
\vskip -6.5cm \hspace*{7cm}
\includegraphics[width=6.5cm]{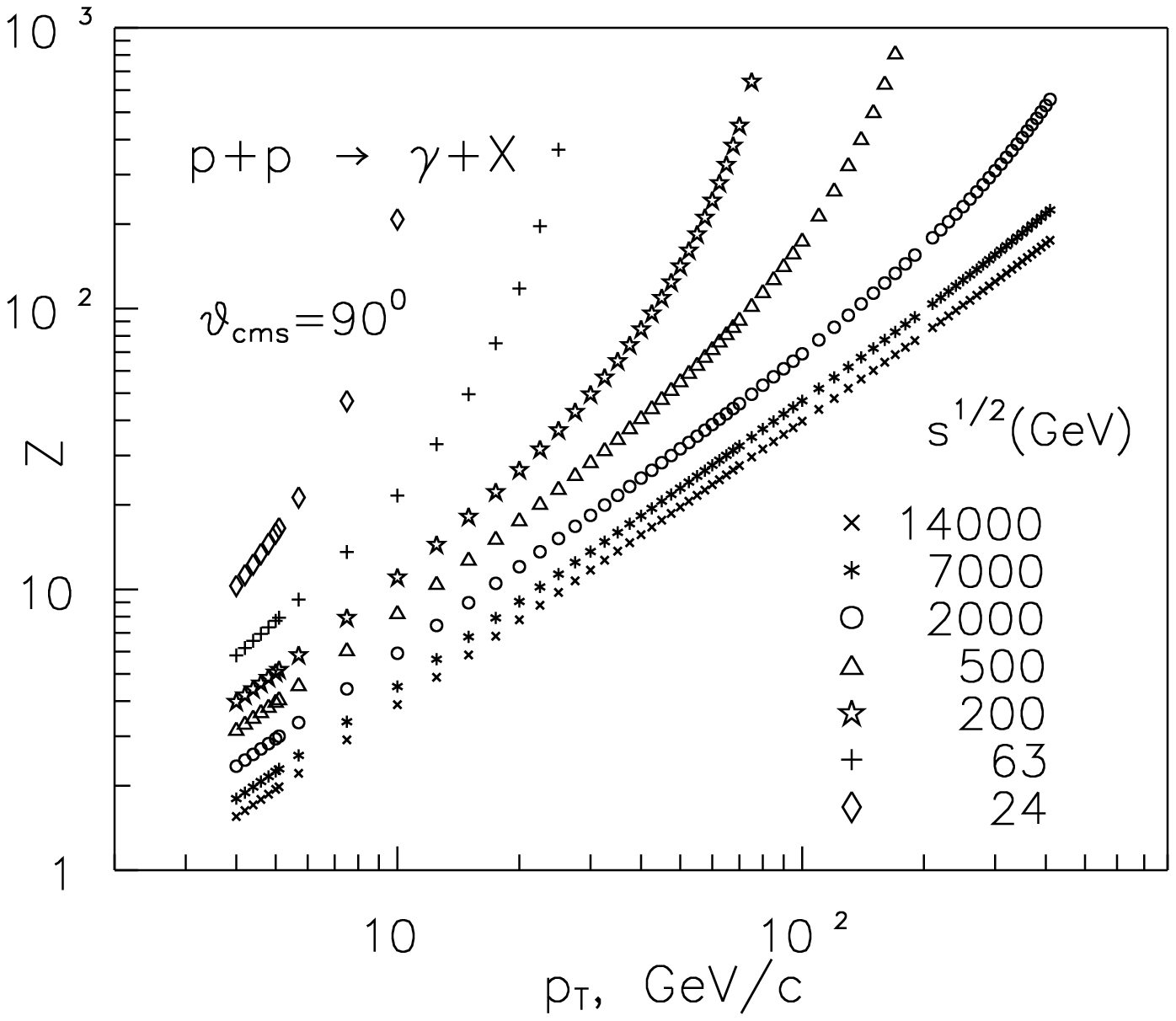}

\hspace*{1cm} c) \hspace*{7cm} d) \caption{
 Dependence of inclusive cross sections of direct photon (a),
$\pi^0$-meson (b) and $\eta^0$-meson (c)  production on
transverse momentum in $pp$ collisions at $\sqrt s =
24-14000~GeV$. Experimental data are taken from
\cite{R806,UA6p,Angel,Kourk}. Solid lines and points   $\star,
\triangle, *, \times$ are the calculated results. (d) Dependence
of the variable $z$ of direct photons produced in $pp$ collisions
on transverse momentum $p_{T}$ at energy $\sqrt s = 24-14000~GeV$
and $\theta_{cms} \simeq 90^0$. }
\end{center}
\end{figure}

{\subsection{$pA$ collisions }}

It is assumed that direct photons  produced in heavy ion
collisions at RHIC and LHC could give a direct indication of
phase transition to the new state of nuclear matter, QGP.
%The modification of photon $p_T$-spectra could be qualitative criterion
%of the transition.

Taking into account an experimental accuracy of data used in the
analysis, the obtained results show that the fractal dimension
$\delta$ and  the slope parameter $\beta$ are independent of A.

%Direct $\gamma$ is considered to be
%a good probe to study the QGP formation.
%Therefore high-$p_T$ direct photon spectra should be
%sensitive to the transition.

Figure 7 demonstrates our predictions of the dependence of the
inclusive cross section  $Ed^3\sigma /dq^3$ on transverse
momentum  $p_T$  for direct photon (a), $\pi^0$ (b) and $\eta^0$
(c) in $pPb$ collisions  at RHIC and LHC energies and at the angle
of $\theta_{cms}=90^0$.

A change of the shape of photon $p_T$ spectra  means  a
modification of the mechanism of photon formation in transition
region.

\begin{figure}
\begin{center}
\hspace*{-8cm}
\includegraphics[width=6.5cm]{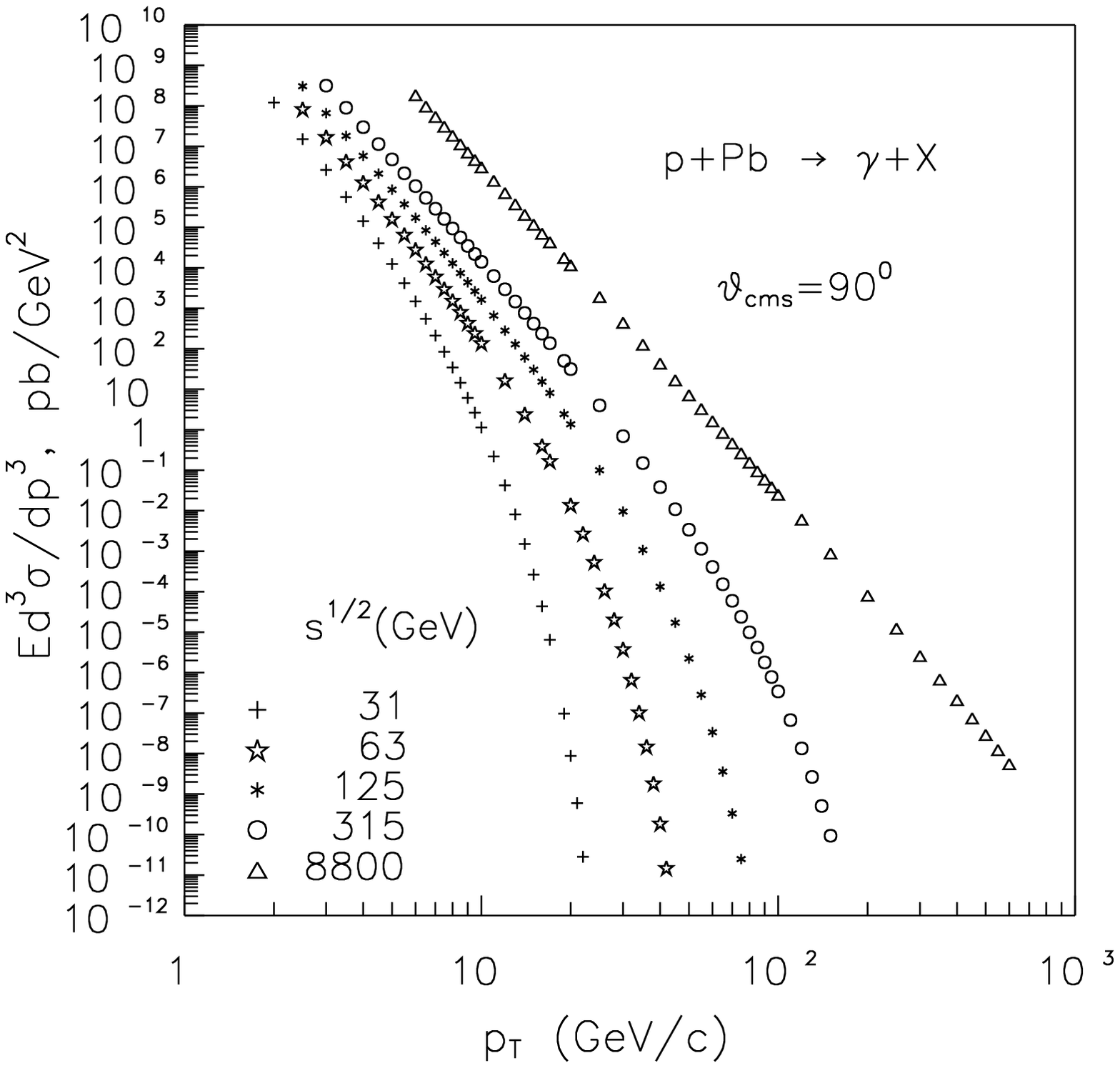}
\vskip -6.cm \hspace*{7cm}
\includegraphics[width=6.5cm]{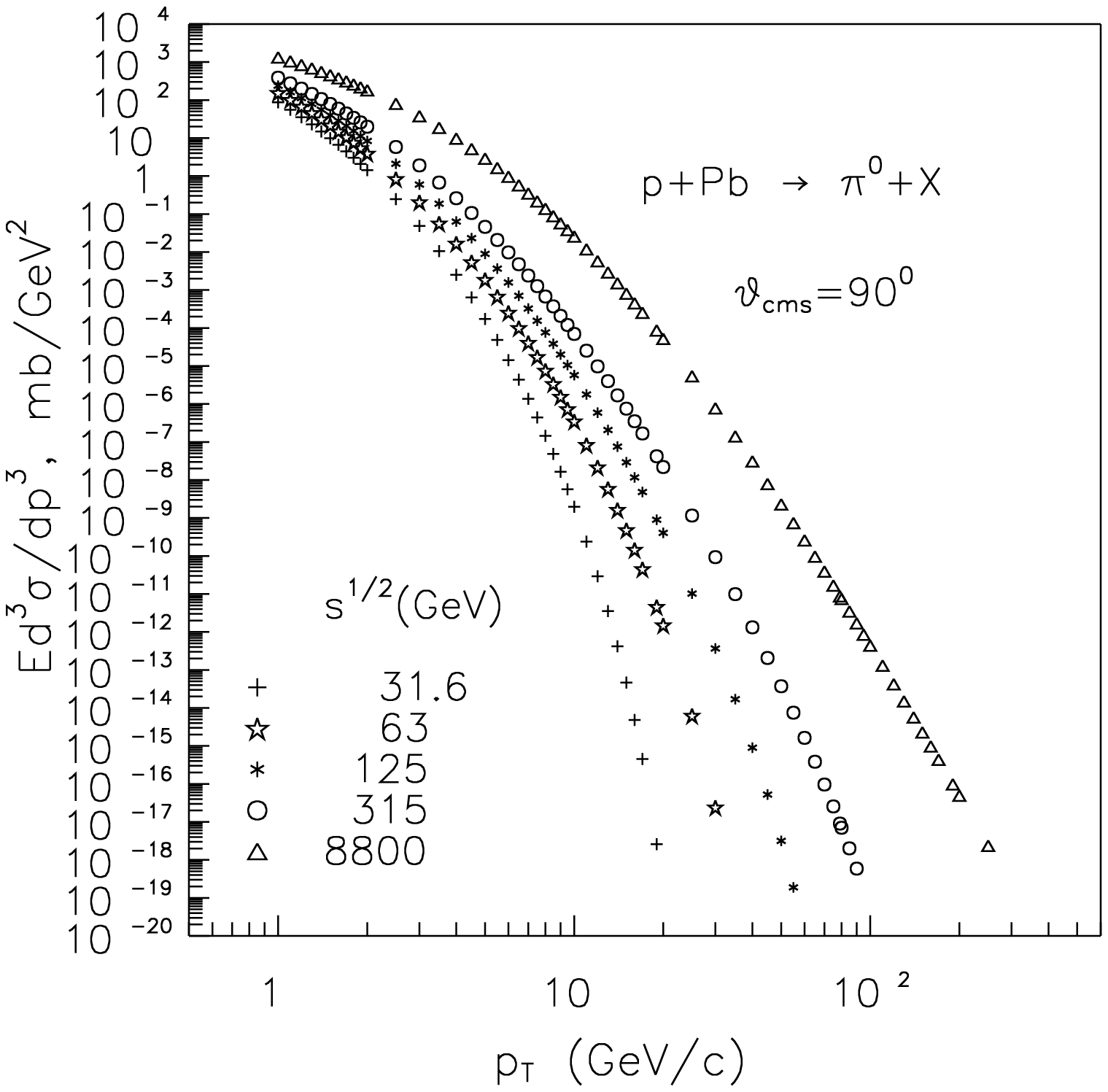}

%\caption{The CERN $\bar{\rm p}$ complex.}
\vskip 0.5cm

\hspace*{1cm} a) \hspace*{7cm} b)
\end{center}
%\end{figure}

%\begin{figure}
\begin{center}
%\hspace*{-8cm}
\includegraphics[width=6.5cm]{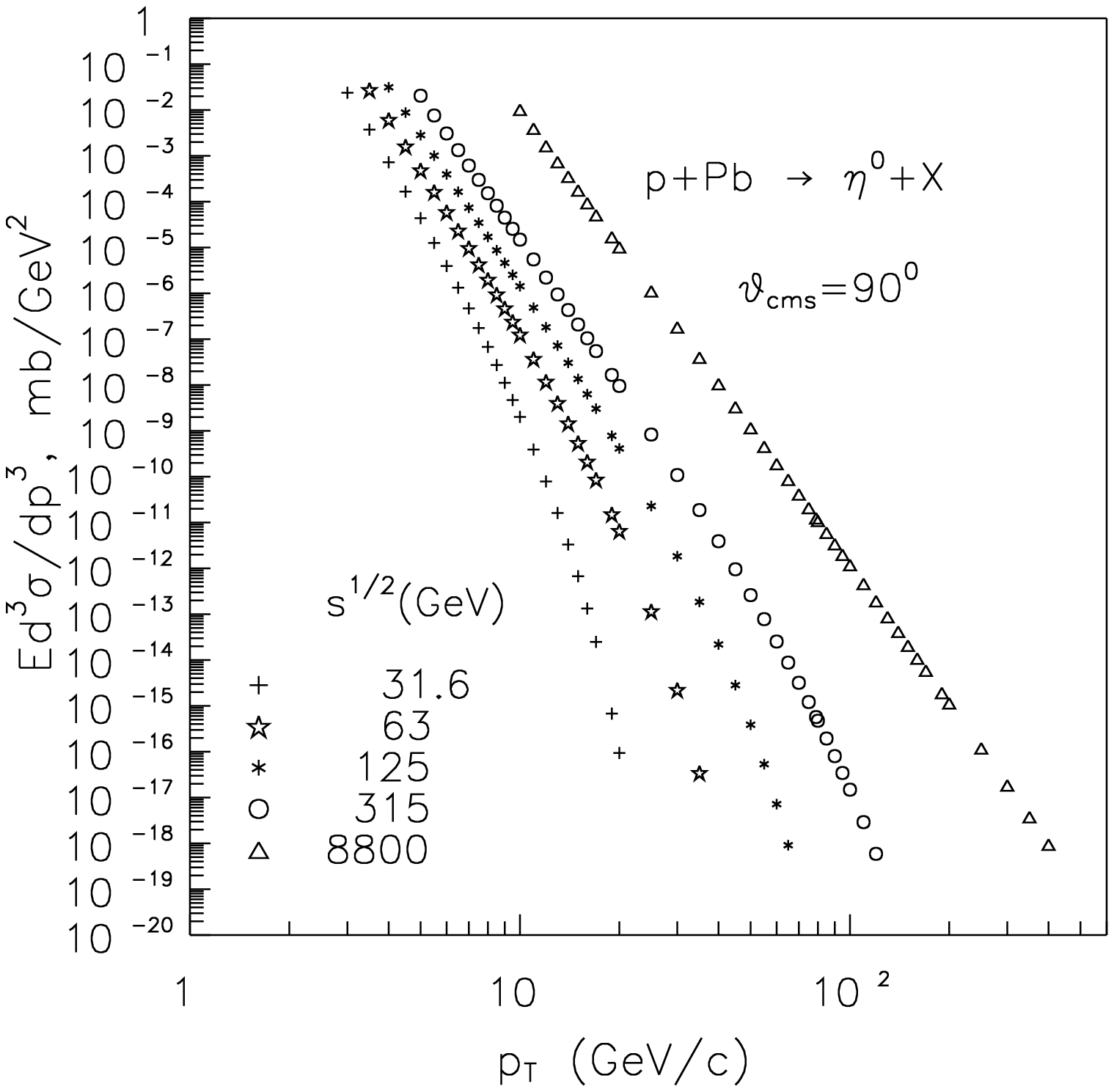}
%\vskip -6.5cm
%\hspace*{7cm}
%\includegraphics[width=6.5cm]{zppg_pt.eps}

\hspace*{1cm} c)
%\hspace*{7cm} d)
 \caption{ Dependence of
inclusive cross sections of direct photon (a), $\pi^0$-meson (b)
and $\eta^0$-meson (c)  production on transverse momentum in
$pPb$ collisions at $\sqrt s = 31-8800~GeV$. Points ($\triangle,
\circ, *, \star, +$ ) are the calculated results.}

\end{center}
\end{figure}

{\section{$z-p_T$  PLOT}}

The $z-p_T$ plot allows us to determine the high transverse
momentum range interesting for searching for the kinematic region
where the $z$-scaling can be violated.
 Figure 6(d) shows the  $z-p_{T}$ plot for the $pp\rightarrow \gamma X$
 process  at $\sqrt s = 24-14000~GeV$.
As seen from Figure 3(b) the scaling function $\psi(z)$ is
measured up to $z\simeq 20$. The function $\psi(z)$ demonstrates
the power behavior
 in the range.
Therefore the  kinematic range $ z > 20$ is of more preferable for experimental investigations of $z$-scaling
violation. The condition determines the low boundaries for $p_T$ ranges,
  $p_T > 5, 10, 16, 22, 35, 45$ and
$52~GeV/c$ at different energy  $\sqrt s = 24, 63, 200, 500, 2000, 7000$ and $14000~GeV$, respectively.

{\section{CONCLUSION}}

Analysis of the numerous experimental data on high-$p_T$ direct photon and $\pi^0$-meson
 production in $pp, \bar pp$ and $pA$ collisions
 obtained at ISR, SpS and Tevatron in the framework of $z$-scaling concept
 was presented.
  It was shown that the general concept of $z$-scaling is
 valid for photon production in hadron-hadron and
 hadron-nucleus collisions.

  The scaling function $\psi(z)$ and scaling variable $z$
  are expressed via the experimental quantities, momenta  and  masses of colliding  and produced particles
  and the invariant inclusive cross section $Ed^3\sigma/dq^3$  and the multiplicity  density of charged
  particles $\rho_A(s,\eta)$.
 The physics interpretation of the scaling function $\psi$ as a
 probability density to produce a particle with the formation length $z$
 is argued. The quantity  $z$ has the property of the fractal measure
 and $\delta $  is the anomalous fractal dimension describing the
 intrinsic structure of the interaction constituents revealed
 at high energies. The fractal dimensions of nuclei satisfy the relation $\delta_{A} = A\cdot \delta_N$.

 It was shown that the properties of $z$-scaling,
 the energy and angular independence, the power law
 $\psi(z)\sim z^{-\beta}$ and  $A$-dependence
 are confirmed by the numerous experimental data obtained at
 ISR, SpS and Tevatron.

 A comparison of the scaling function of direct photon and $\pi^0$-meson production in $pp$ and $\bar pp$
 was performed and  different asymptotic behavior of $\psi(z)$  was found.  It was shown that
 $p_T$-dependence of the ratio of direct photon and $\pi^0$-meson inclusive cross sections for
 $pp$ and $\bar pp$ collisions has the different crossover points.
 Based on the universality of the  scaling function, the predictions  of direct photon, $\pi^0$- and
 $\eta^0$-meson cross sections in $pp$ and $pPb$ collisions at RHIC and LHC energies were made. The $z-p_T$
 plot was used to establish the kinematic range that is of more preferable for experimental investigations of
 $z$-scaling violation.

The violation of $z$-scaling due to the change of the value of the fractal dimension $\delta$ is suggested to
search for a new physics phenomena such as quark compositeness, new type of interactions, nuclear phase
transition in $pp, pA$ and $AA$ collisions at  RHIC and LHC.

\noindent

\section*{ACKNOWLEDGEMENTS}

One of the authors (M.T.) would like to thank the organizers  of the workshop "Hard Probes in Heavy Ion
Collisions at LHC" U. Wiedemann, H. Satz,  M. Mangano and convenors of the working groups P. Aurenche, P.
Levai and K.J. Eskola  for excellent atmosphere during the workshop and for many interesting and useful
discussions.

\end{document}